\documentclass[final,3p,times,twocolumn]{elsarticle}

\usepackage{graphics}
\usepackage{graphicx}
\usepackage{epsfig}

\usepackage{amssymb}
\usepackage{amsthm}

\usepackage[normalem]{ulem}
\usepackage{color}
\definecolor{blue}{rgb}{0,0,1}
\definecolor{red}{rgb}{1,0,0}

\usepackage{lineno}

\biboptions{comma, square}

\journal{Nuclear Instruments and Methods in Physics Research A}

\begin{document}

\begin{frontmatter}



\title{Design and Performance of the Data Acquisition System for the NA61/SHINE Experiment at CERN\tnoteref{titlefootnote}}
\tnotetext[titlefootnote]{This paper is dedicated to the memory of our collegue, Dr.\ Ervin D\'enes, coauthor of this paper, deceased in June 2015.}


\author[wigner]{Andr\'as L\'aszl\'o\corref{corlabel}}
\ead{laszlo.andras@wigner.mta.hu}
\cortext[corlabel]{Corresponding author.}
\author[wigner]{Ervin D\'enes\fnref{ervinfootnote}}
\fntext[ervinfootnote]{Deceased.}
\author[wigner]{Zolt\'an Fodor}
\ead{fodor.zoltan@wigner.mta.hu}
\author[wigner]{Tivadar Kiss}
\ead{kiss.tivadar@wigner.mta.hu}
\author[uca]{Stuart Kleinfelder}
\ead{stuartk@uci.edu}
\author[cern]{Csaba So\'os}
\ead{csaba.soos@cern.ch}
\author[warsaw]{Dariusz Tefelski}
\ead{dariusz.tefelski@cern.ch}
\author[wigner]{Tam\'as T\"olyhi}
\ead{tolyhi.tamas@wigner.mta.hu}
\author[wigner]{Gy\"orgy Vesztergombi}
\ead{vesztergombi.gyorgy@wigner.mta.hu}
\author[cern,cracow]{Oskar Wyszy\'{n}ski}
\ead{oskar.wyszynski@cern.ch}
\address[wigner]{Wigner Research Centre for Physics of the Hungarian Academy of Sciences, P.O.Box 49 H-1525 Budapest, Hungary}
\address[uca]{University of California, Irvine, CA 92697-2625, USA}
\address[cern]{CERN, CH-1211 Geneva 23, Switzerland}
\address[warsaw]{Faculty of Physics, Warsaw University of Technology, Koszykowa 75, 00-662 Warsaw, Poland}
\address[cracow]{Faculty of Physics, Dep.\ of Astro.\ and Appl.\ Comp.\ Sci.\ of the Jagiellonian University, Stanis\l{}awa \L{}ojasiewicza 11, 30-348 Krak\'ow, Poland}

\begin{abstract}
This paper describes the hardware, firmware and software systems used in data 
acquisition for the NA61/SHINE experiment at the CERN SPS accelerator. 
Special emphasis is given to the design parameters of the readout electronics 
for the 40$\,\mathrm{m}^{3}$ volume Time Projection Chamber detectors, as these give the 
largest contribution to event data 
among all the subdetectors: events consisting of 8$\,\mathrm{bit}$ ADC values from 256 time 
slices of 200$\,\mathrm{k}$ electronic channels are to be read out with $\sim$100$\,\mathrm{Hz}$ 
rate. The data acquisition system is organized in ``push-data mode'', 
i.e.\ local systems transmit data asynchronously. Techniques of solving 
subevent synchronization are also discussed.
\end{abstract}

\begin{keyword}
DAQ \sep readout \sep TPC \sep NA61 \sep SHINE \sep SPS \sep CERN \sep DDL \sep RORC
\PACS 07.05.Hd
\end{keyword}

\end{frontmatter}


\section{Introduction}
\label{introduction}

NA61/SHINE is a large acceptance fixed target hadron spectrometer experiment 
at the SPS accelerator at CERN \cite{na61homepage, na61proposal, na61cpod07, na61qm09, na61nim}. 
The main tracking system, i.e.\ the bending magnets and the five TPC detectors 
are inherited from the former experiment NA49 \cite{na49detector}. 
The physics program of NA61/SHINE is quite complex and covers the search 
for the critical point of strongly interacting matter, 
study the onset of deconfinement, the quantification of medium effects in strongly 
interacting matter, furthermore the measurement of hadron 
production spectra in hadron-nucleus collisions for cosmic ray and neutrino 
physics applications. These studies are carried out using the SPS beamline which is 
able to provide hadron beams in the 10-350$\,\mathrm{GeV/c}$, and ion beams in the 
10-160$\,\mathrm{GeV/c/nucleon}$ beam momentum range. The hadron beams are produced by the 
400$\,\mathrm{GeV/c}$ proton beam of the SPS accelerator hitting a beryllium production target and are 
tagged with their particle type using a differential \v{C}erenkov detector trigger signal. 
Ion beams are either direct beams from SPS whenever 
compatible with the accelerator schedule, or are fragmented from a lead beam of 
the SPS using a beryllium fragmentation target with subsequent 
element tagging using a threshold \v{C}erenkov detector or via scintillator response amplitude. 
The beam rate reaching the detector is up to 100$\,\mathrm{kHz}$, being 
the safety limit, out of which about 5 to 100\% are the tagged useful beam 
particles with the selected type, depending on the actually used configuration. 
This beam hits a fixed target 
which is either liquid hydrogen or solid state material, depending on the 
reaction to be studied. The thickness of the target is adjusted in such a way 
that the collision probability of the selected beam particle type with the 
target material is around 0.1-3\% in order to limit the contribution of 
secondary collisions within the target material. This setting provides 
$\sim$1-3$\,\mathrm{kHz}$ collision event rate of the right type to be potentially recorded by 
the spectrometer downstream of the target.

\begin{figure}[!ht]
\begin{center}
\includegraphics[width=7.7cm]{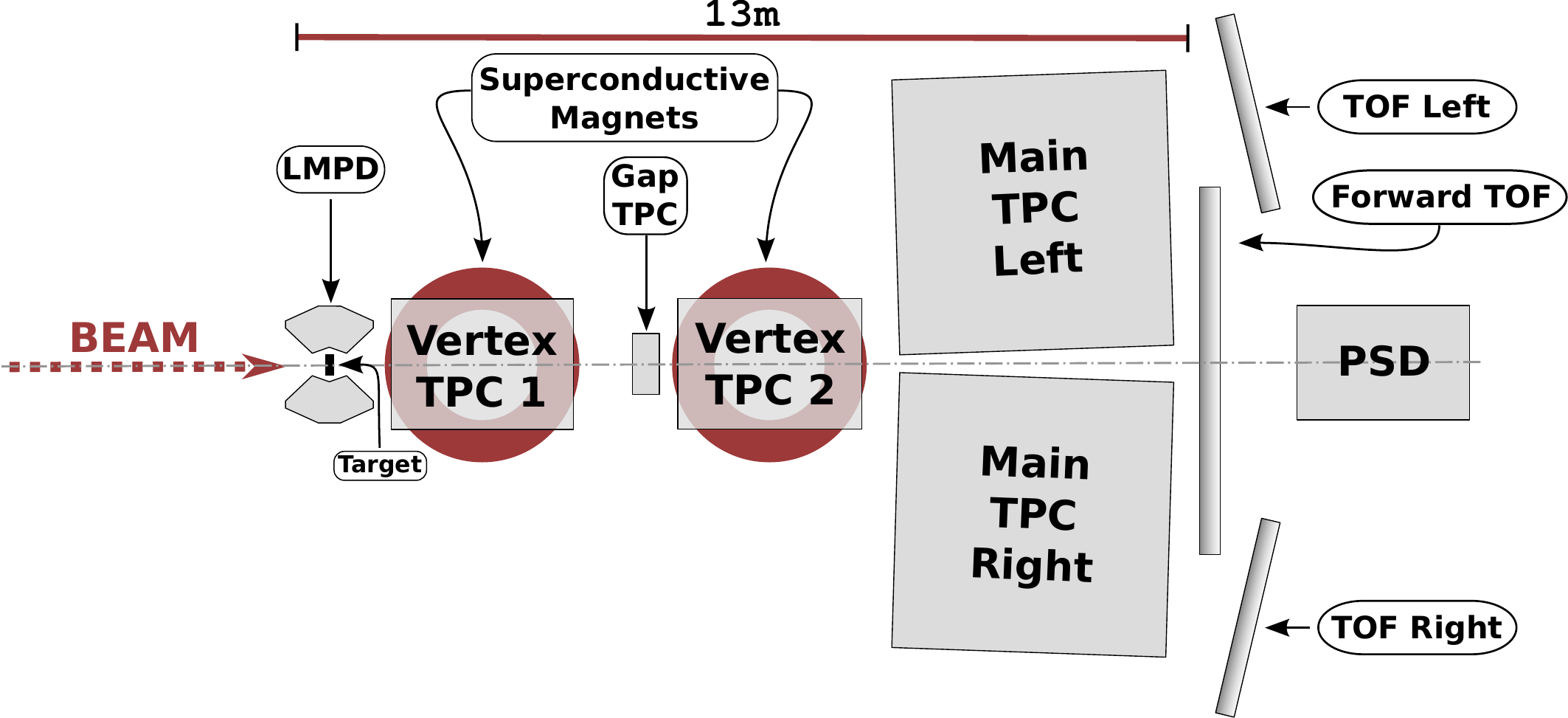}
\caption{The setup of the NA61/SHINE detector. 
A set of trigger and beam counters are placed on the beamline, followed 
by the fixed target. A large acceptance magnetic spectrometer setup with 
40$\,\mathrm{m}^{3}$ tracking volume is placed downstream of the target. 
The spectrometer setup consists of two large superconducting magnets and 
TPC (Time Projection Chamber) volumes. The setting is completed by a set of 
ToF (Time-of-Flight detectors) along with a calorimeter in the beamline, called PSD 
(Projectile Spectator Detector). 
In some of the runs the target is surrounded by a special detector, called 
LMPD (Low Momentum Particle Detector) \cite{lmpd}, which is a small TPC 
chamber using the same readout as the large ones.}
\label{na61outline}
\end{center}
\end{figure}

The outline of the experimental setup is shown in Fig.~\ref{na61outline}. The ultrarelativistic particles 
produced in the collision within the target enter into the strong, $\sim$0.1-1.5$\,\mathrm{Tesla}$, field of two 
superconducting bending magnets where the trajectories 
are deflected according to their momentum to charge ratio. The bending 
power of the magnet system is up to 9$\,\mathrm{Tm}$. 
Within and downstream of the bending magnets, five large volume TPC 
(Time Projection Chamber) detectors record the charged 
particle trajectories in a 40$\,\mathrm{m}^{3}$ tracking volume. 
A set of ToF (Time of Flight) detectors aid the 
particle identification. The most downstream detector of the experiment is 
the PSD (Projectile Spectator Detector), a calorimeter used to measure the energy fraction of 
the piece of projectile nucleus which did not take part in particle production, 
providing a geometric measure of collision centrality.

Upon the start of the NA61/SHINE program, several new detectors were added to 
the original NA49 tracking setup. In addition, the available readout rate 
not greater than $\sim$10$\,\mathrm{Hz}$ in NA49 was insufficient for fully exploiting the available 
beam rates, and was also insufficient for completing the data taking schedule 
in a timely manner with the available 3-5 month of beamtime per year. 
Therefore a decision was made to design and produce a new readout system for 
NA61/SHINE in order to be able to record events at a rate $\sim$100$\,\mathrm{Hz}$, 
along with incorporating new detectors into the readout chain. In this upgrade 
project the most elaborate part was a new solution for the TPC readout, as this detector 
component gives the largest contribution to the raw data. Therefore, this paper gives a 
special emphasis on the design requirements and description of the TPC readout 
part of the NA61 DAQ (Data AcQuisition) system.

The paper is organized as follows. Section~\ref{tpcrequirements} summarizes 
the requirements on the new TPC electronics as a main motivation. 
Section~\ref{tpcelectronics} describes the new TPC electronics, their firmware 
and working principle. 
Section~\ref{psdcamactofelectronics} describes the integration of further 
subdetectors into the readout chain. Section~\ref{dataflow} summarizes the 
signal and data flow lines within the experimental setup.
Section~\ref{synchronization} 
describes the solutions for guaranteeing the event synchronization in the 
parallel data channels. Section~\ref{onlinesoftware} describes the 
online software, mainly the Central DAQ. Section~\ref{rawdatastructure} 
outlines the structure of the raw data files written by Central DAQ. 
Section~\ref{performance} summarizes 
the observed performance. Section~\ref{concludingremarks} concludes the paper.

\section{Requirements on the TPC readout electronics}
\label{tpcrequirements}

The 40$\,\mathrm{m}^{3}$ TPC chamber system produces the largest and most complicated subevent data of the 
experiment: for each triggered event the charge deposit on each of the 182784 
readout pads need to be acquired throughout the 51.2$\,\mathrm{\mu sec}$ drift time in 
256 or 512 samples, where the charge deposit is measured in 1$\,\mathrm{Byte}$ ADC counts. 
This means 50 or 100$\,\mathrm{MByte}$ per event in the 256 or 512 
time sampling mode, respectively. In normal data taking, the 256 timebin mode is used, the 
512 timebin mode is needed only during short and rare periods of gain 
calibration runs with radioactive Krypton gas in the TPC volumes. During normal 
beam data taking 50$\,\mathrm{MByte}$ per event needs to be transported 
at a $\sim$100$\,\mathrm{Hz}$ event rate, which would result in $\sim$5$\,\mathrm{GByte/sec}$ data rate. 
However, the TPC detector data is zero 
dominated, since detection signal in the TPC volume only appears where 
particle trajectories traverse it. Therefore, it is evident that some kind 
of data compression method can and should be applied.

The original NA49 FEE (Front-End Electronics) cards of the TPC 
chambers \cite{na49readout, na49fesca} provided a minimal dead time of 
11.7$\,\mathrm{msec}$ when read out. This already allows 86$\,\mathrm{Hz}$ readout rate without 
modification of the 5712 pieces of FEE cards. A trade-off between the needed 
readout rate and the development time and costs motivated the re-usage of the 
NA49 FEE cards in the new system without modification, and the redesign of the 
readout electronics upstream of them. The electronic units reading out the 
FEE cards are called MotherBoards. They perform the steering of the readout 
process of the FEEs, the data compression before transfer, and subsequent 
serial transmission of the processed FEE data. Because of geometrical and 
data density constraints, the MotherBoards were designed to read out 
up to 24 FEE cards, and therefore 248 MotherBoards were used in the complete 
system.

Due to the extended size of the full experimental setup, event data need to 
be transmitted to large distances, about 50$\,\mathrm{m}$, from the detectors to the 
location of the Central DAQ in the control room. This means a high risk for accidental 
introduction of ground loops in the system, and therefore galvanically decoupled 
transmission lines were needed. Our choice fell to the relatively cheap and 
well understood large bandwidth DDL (Detector Data Link) system \cite{ddl, rorc} for long range optical 
data transfer, also used in the ALICE experiment working at the LHC accelerator 
at CERN.

In order to minimize the number of optical links toward Central DAQ, an 
intermediate serialization stage was needed in between the MotherBoards 
and the Central DAQ. These units were called the ConcentratorBoxes and were 
designed to serialize data of up to 32 MotherBoards onto a DDL line. Short 
distance data transfer between the MotherBoards and the ConcentratorBoxes 
used relatively inexpensive LVDS (Low Voltage Differential Signal) 
connections. These connections, 
being differential, are noise tolerant and although they do not 
provide galvanic isolation, can work with up to $\pm$1$\,\mathrm{V}$ common mode 
mismatch between the transmitter and the receiver side.

\section{The TPC readout electronics}
\label{tpcelectronics}

\begin{figure}[!ht]
\begin{center}
\includegraphics[width=7.7cm]{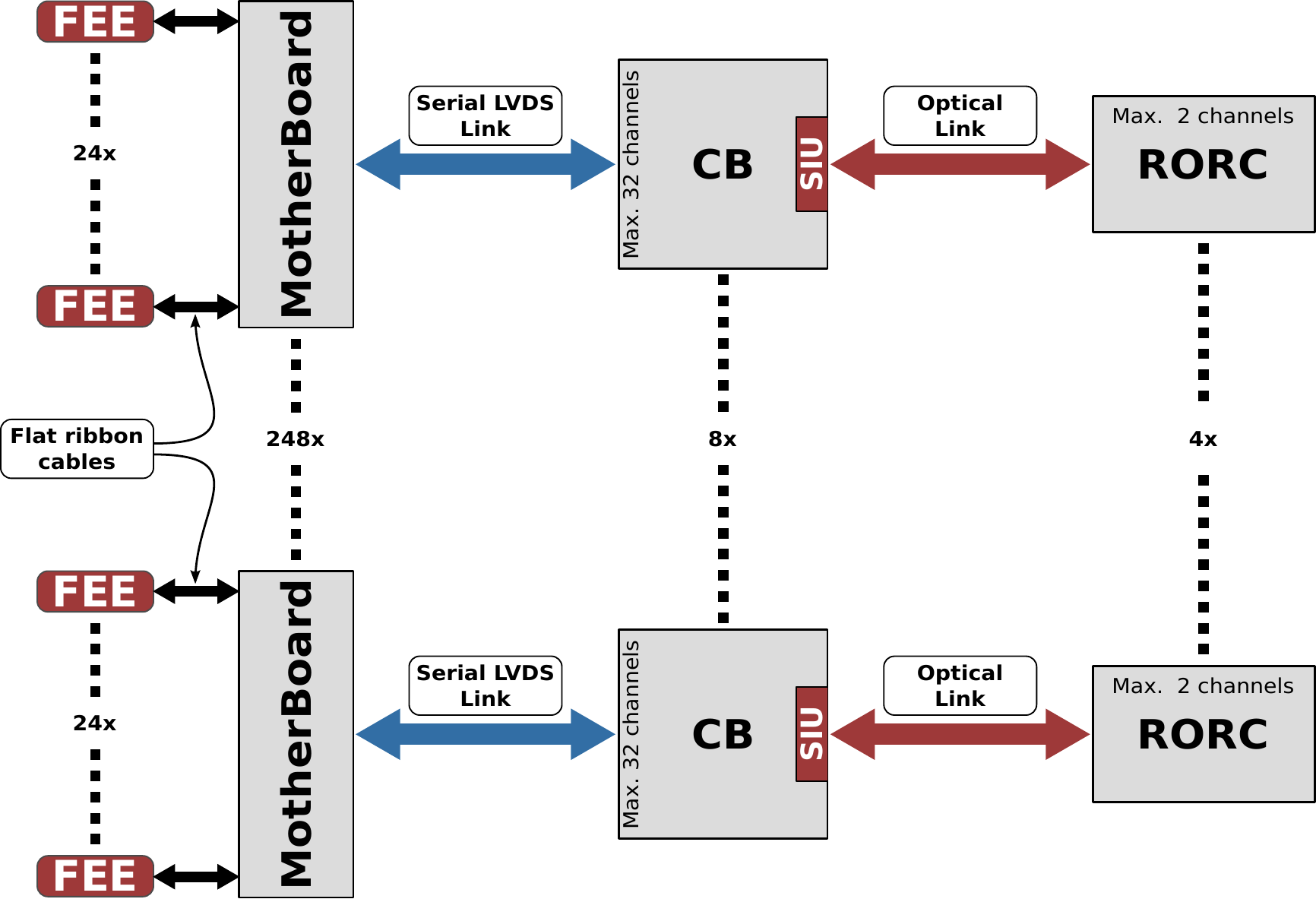}
\caption{Block diagram of the NA61 TPC readout system. Up to 24 FEE cards are 
accessed by a processing MotherBoard using flat cable connection, then up to 
32 MotherBoards transmit their data to a ConcentratorBox using LVDS connection, 
finally up to 10 ConcentratorBoxes forward the data to the Central DAQ PC using 
DDL optical line.}
\label{tpcreadoutscheme}
\end{center}
\end{figure}

Motivated by the requirements discussed in Sec.~\ref{tpcrequirements}, 
the NA61 TPC readout system \cite{specs} has four main components as shown in 
Fig.~\ref{tpcreadoutscheme}.

The main parts of the assembly are:
\begin{enumerate}
\item  Front-End Electronics \cite{na49readout}: these cards perform the analog sampling of TPC 
       pad charges, the analog storage of these and subsequent digitalization. 
       Upon trigger arrival, one FEE card samples the preamplified and shaped 
       pad charges over 51.2$\,\mathrm{\mu sec}$ time duration in 256 or 512 
       timeslices using an SCA (Switched Capacitor Array) \cite{na49fesca}. The 
       pertinent sampling clock is derived from a centrally generated 25$\,\mathrm{MHz}$ 
       oscillator in order to avoid phase ambiguity. 
       Following the sampling period the, stored charges of each timeslice are digitized to 9$\,\mathrm{bit}$s 
       using Wilkinson ADCs. One FEE can handle 32 TPC detection pads. 
       Due to historical reasons, each FEE is built up of two equivalent 
       halves each handling 16 pads. The FEE cards are controlled and read out 
       via a 50 line flat cable by a MotherBoard.
 \item MotherBoards \cite{specs}: these organize the readout of the FEE cards and build 
       a subevent data structure. They also perform pedestal subtraction, 
       noise suppression, bit truncation and zero compression\footnote{A 
       possibility for using Huffman compression is also foreseen, but it 
       is not yet implemented, as the compression rate of simple zero compression 
       proved to be sufficient for present data taking settings.}. 
       A MotherBoard has 24 input ports which are capable 
       of reading out FEEs in parallel.
       For the 24 FEE cards two different cabling topologies are possible: 
       one FEE card on each port (24$\times$1 cabling), and 2 FEE cards connected in 
       daisy chain by a flat cable to 12 ports of a MotherBoard (12$\times$2 cabling). 
       After reading out the FEE cards, the ADC values are 
       pedestal subtracted, and any residual values smaller than a typical threshold of 
       6$\,\mathrm{ADC}$ ($\sim$2$\,\sigma$ of the noise level) are 
       substituted with zero values in a noise suppression process. These 
       pedestal subtracted and noise suppressed ADC values are then truncated to 8$\,\mathrm{bit}$s, 
       are ordered according to natural detector geometry, and are 
       subsequently serialized into a data stream. This 
       data stream is then zero compressed, i.e.\ in case of zero ADCs only one 
       zero and the number of consecutive zeros (up to 255) are transmitted in 
       the stream. The serialized data of the FEEs are then sent by the 
       MotherBoard onto an 50$\,\mathrm{Mbit/sec}$ effective bandwidth LVDS connection, its transmission medium 
       being a maximum 15$\,\mathrm{m}$ long STP (Shielded Twisted Pair) cable. 
       The LVDS connection reaches the next stage, the ConcentratorBoxes.
 \item ConcentratorBoxes \cite{specs}: these are standalone serializer/deserializer 
       boxes with 32 bidirectional LVDS inputs/outputs. 
       A maximum of 32 MotherBoards can be connected to one ConcentratorBox. During the 
       readout, the ConcentratorBox serializes the incoming data 
       streams onto a single 32$\,\mathrm{bit}$ wide data stream.
       Then the 32$\,\mathrm{bit}$ data is transmitted to the DAQ server PC through DDL connection.
       In the other directions, the ConcentratorBox receives commands, 
       pedestal table and other data from the DAQ PC addressed to the 
       MotherBoards. The ConcentratorBox routes these data to the 
       selected MotherBoard.
 \item DDL \cite{ddl, rorc}: the DDL links are high-speed, serial optical links 
       with parallel interfaces developed for the ALICE experiment. A link 
       consists of a SIU (Source Interface Unit) card, a PCI-X or PCI-Express RORC (Readout Receiver Card) 
       and a maximum 200$\,\mathrm{m}$ length duplex optical cable with 
       LC connectors. The DDL transfers the acquired detector data, data 
       blocks or status info to the Central DAQ computer, and using a backward 
       channel can load data blocks or commands from the Central DAQ system 
       to the lower level systems. The bandwidth is guaranteed to be a 
       minimum of 200$\,\mathrm{MByte/sec}$ in both directions.
\end{enumerate}

\begin{figure}[!ht]
\begin{center}
\includegraphics[width=7.7cm]{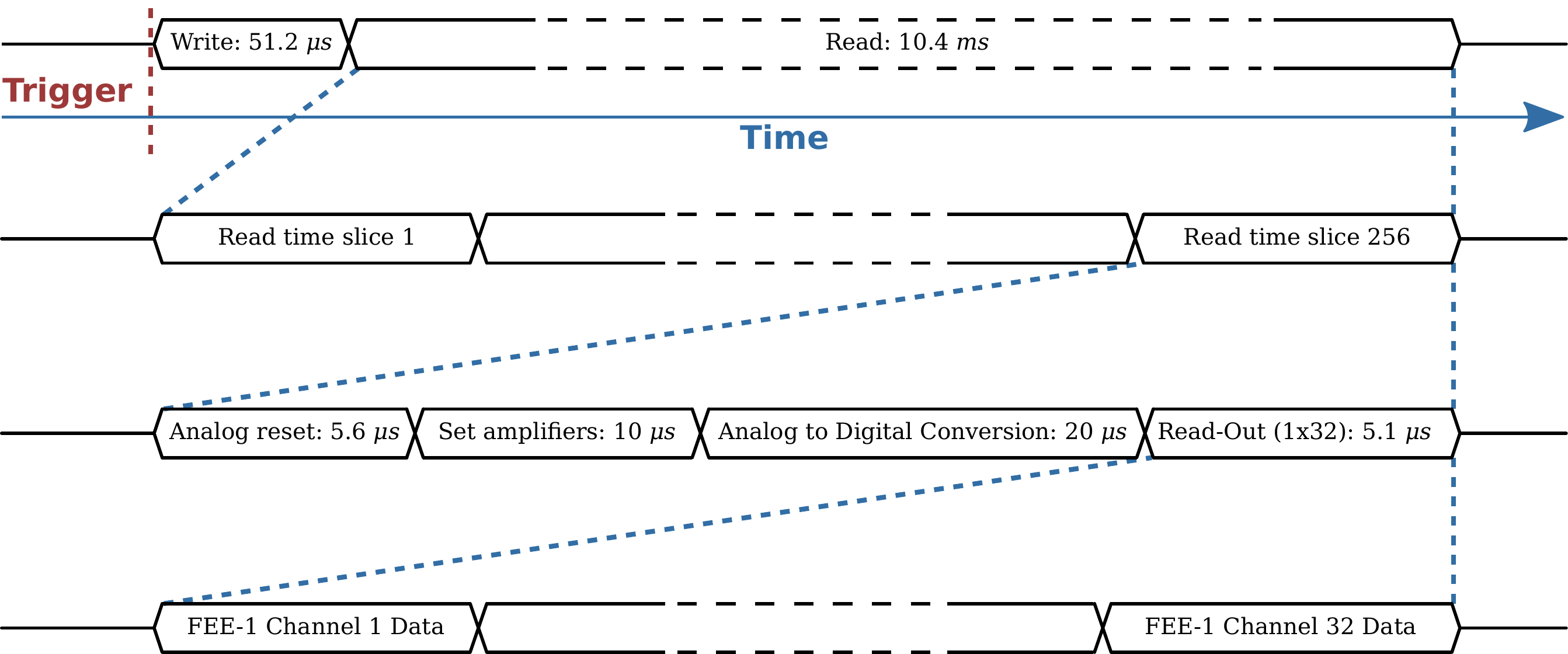}
{\vspace*{10mm}}
\includegraphics[width=7.7cm]{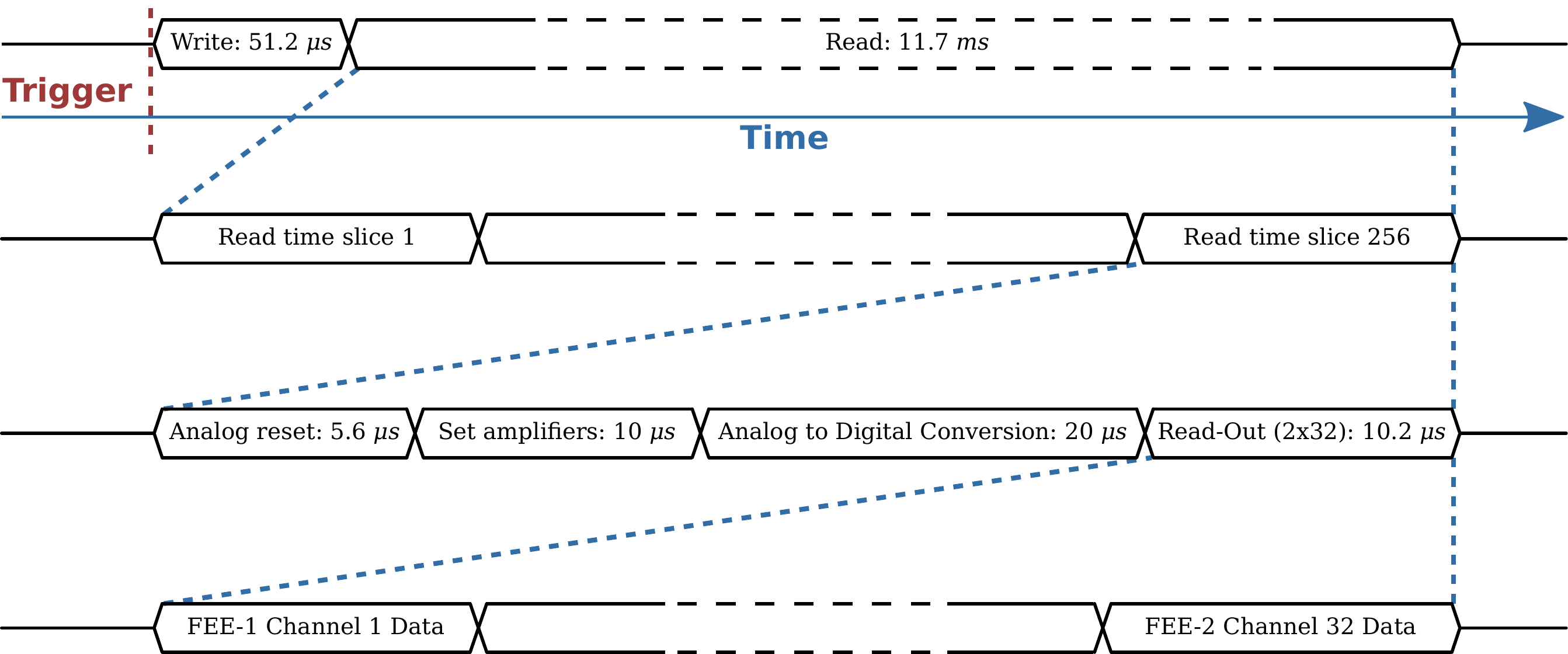}
\caption{Time diagram of readout process of an FEE card in 
24$\times$1 (top panel) and 12$\times$2 (bottom panel) cabling mode. 
The readout is done in the timeslice order, i.e.\ for each timeslice the 
charges on the 1$\times$32 channels of 1 FEE card or 2$\times$32 channels of 
2 FEE card is digitized, and then they are read out in channel order. 
The magnified time regions are indicated by the dashed lines.}
\label{FE_card_read-out}
\end{center}
\end{figure}

The MotherBoard has four different working modes: command processing mode, 
data block uploading or downloading mode, data collecting mode, 
pedestal collecting mode. In case of data collecting or pedestal collecting 
mode, upon receiving a ``Start Data Taking'' command the 
MotherBoard waits for the trigger signals. 
The workflow of the data acquisition is organized as follows. 
A common external clock (TPC CLK) is distributed to the MotherBoards in order 
to synchronize the readout system. The acquisition starts when a common 
external trigger is received. The TPC CLK along with the trigger are distributed via ECL lines. 
Receiving the trigger the MotherBoards force the FEE cards to start 
sampling the analog charges deposited on the TPC pads after amplification and 
shaping. Upon finishing, the MotherBoards issue digitization command and read 
out the raw amplitude ADC data produced on the FEE cards. The digitalization and 
ADC readout are made in timeslice order (Fig.~\ref{FE_card_read-out}):
\begin{enumerate}
\item digitalization of 32 channels in parallel,
\item readout of the 32 channels in channel order for all the FEE-s simultaneously,
\item the above steps are repeated for each timeslice.
\end{enumerate}
The readout time slightly depends on which of the 24$\times$1 or 12$\times$2 
cabling topologies are used for the connection of the FEE cards to the 
MotherBoard, as in the latter case the two FEE card on the same flat cable 
is read out in an alternating order, causing a slight overhead. 
The maximum possible speed using 256 timeslices is 
around 90 events per second, limitation due to the FEE cards 
architecture. The control signals necessary for the FEE readout are 
all produced by the MotherBoard.

\begin{figure}[!ht]
\begin{center}
\includegraphics[width=7.7cm]{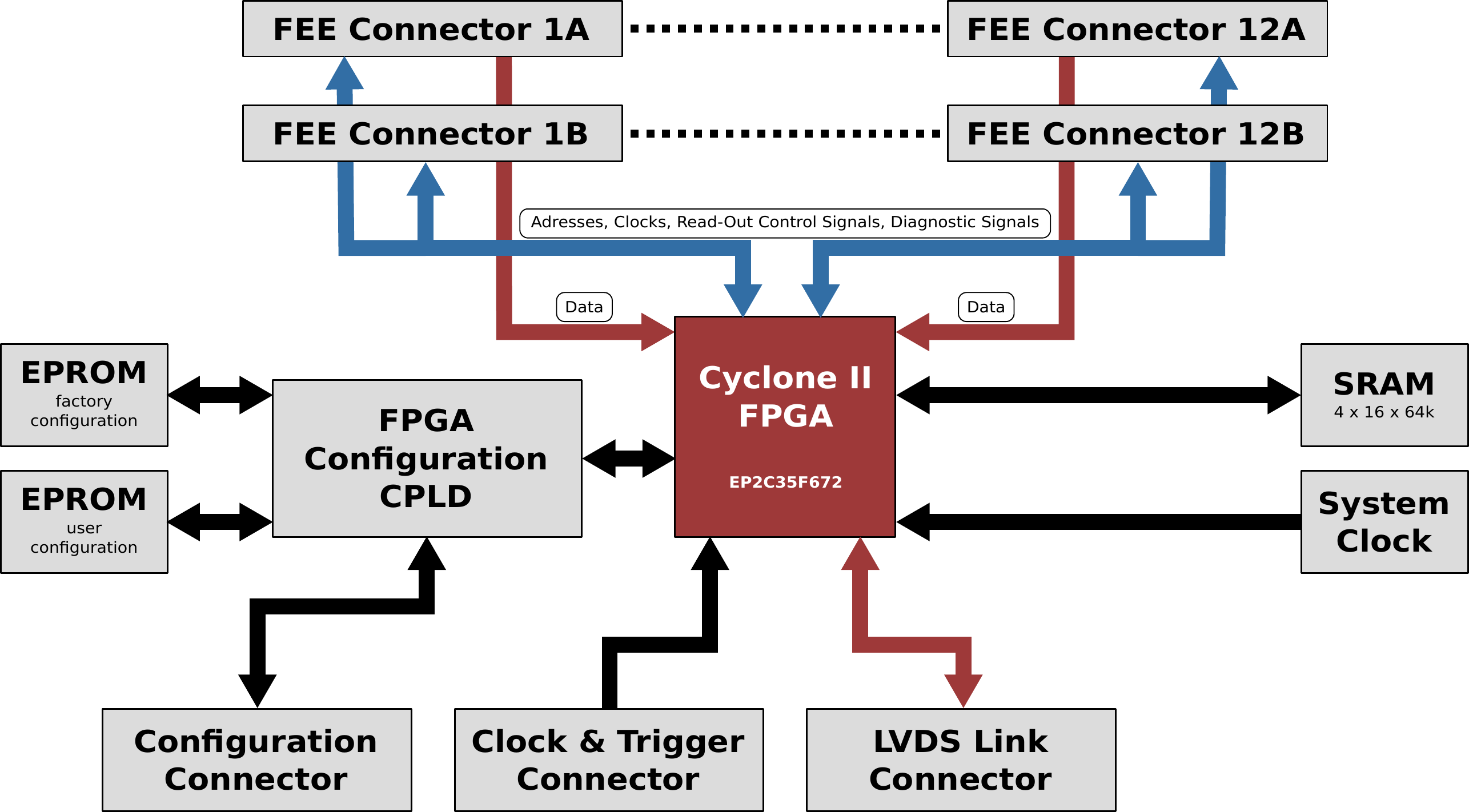}
\caption{Schematic block diagram of a MotherBoard hardware architecture.}
\label{motherboard}
\end{center}
\end{figure}

The schematic block diagram of a MotherBoard hardware architecture, based on 
FPGA technology for maintaining good development flexibility, is seen in 
Fig.~\ref{motherboard}. Each MotherBoard is equipped with 24 flat cable connectors for the 
24 FEE cards. The FEE cards are read out in parallel, and their data are 
temporarily stored in the FPGA. The MotherBoard automatically multiplexes 
the 24 input data stream in a single serial output data stream. 
Before sending the data, the MotherBoard arranges them in padrow order and performs zero 
compression, reducing the amount of data. For the output data, sent on the 
LVDS lines to the ConcentratorBox, a simple, reliable bidirectional link 
protocol has been developed. The pedestal table and the noise suppression 
threshold can be downloaded to the MotherBoard through the same serial link. 
Some parameters important for the readout, such as the TPC and MotherBoard 
location identifiers, or the cabling topology of 24$\times$1 or 12$\times$2 
can be preset using jumpers.

\begin{figure}[!ht]
\begin{center}
\includegraphics[width=7.7cm]{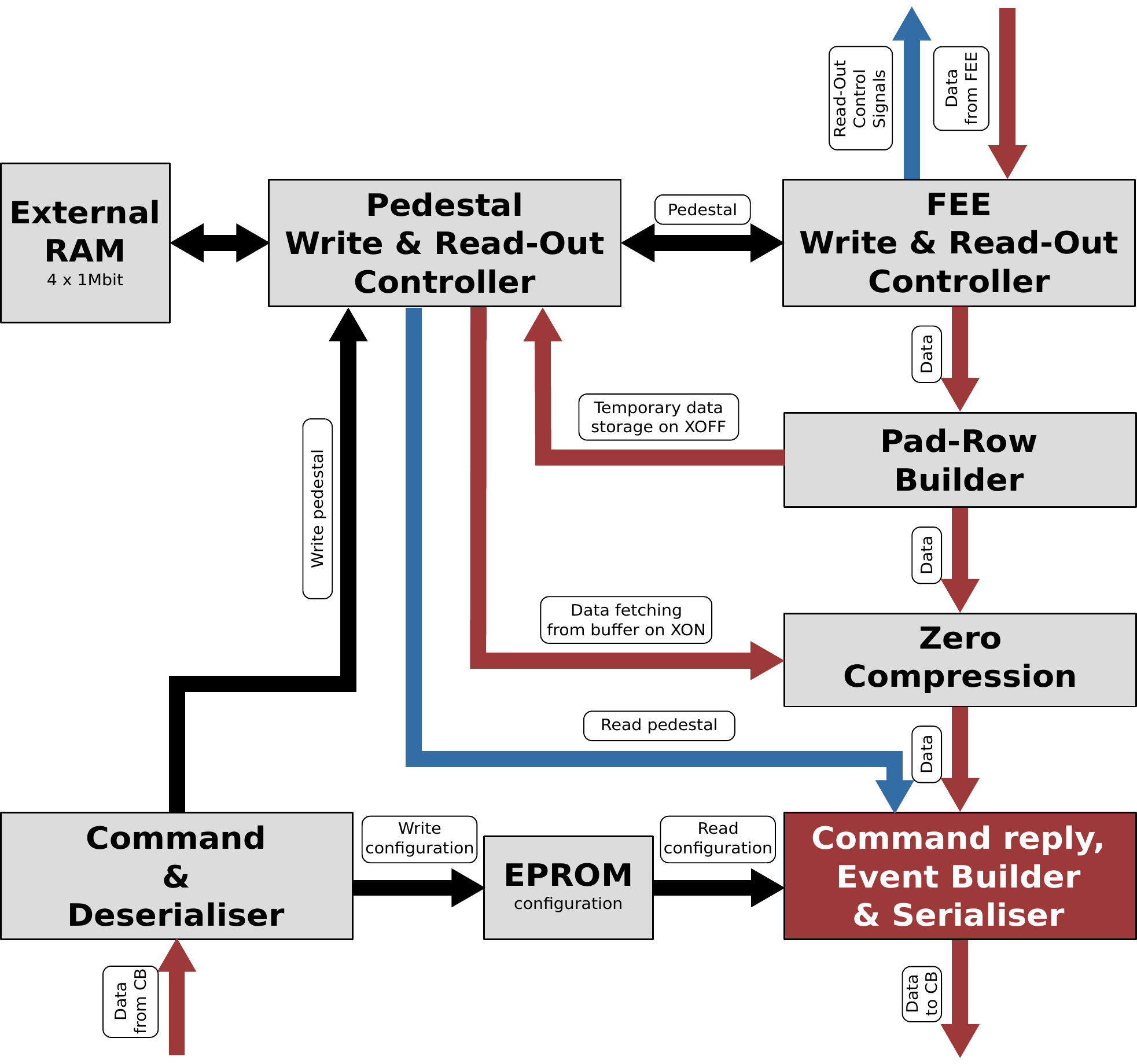}
\caption{Schematic block diagram of the MotherBoard firmware.}
\label{motherboardfw}
\end{center}
\end{figure}

The simplified firmware block diagram of the MotherBoard firmware is 
shown in Fig.~\ref{motherboardfw}.
\begin{itemize}
\item The ``FEE Write and Read-Out Controller'' unit controls the read out of 
the FEE cards and subtracts the pedestal values.
\item The ``Pedestal Write and Read-Out Controller'' module has several 
functions. During data taking it reads the pedestal value from the external 
SRAM memory belonging to the current data. In pedestal collecting 
mode it saves the value in the external SRAM and transmits all pedestals 
toward the ConcentratorBox when the collecting has finished. 
Using the received pedestal, i.e.\ zero-signal events, the DAQ computer calculates the 
table of average pedestals and sends it back to the MotherBoard, saving 
it in its external SRAM memory. This module has an additional function 
which is activated in the rare cases when the output FIFO of the MotherBoard 
becomes full. In that case the incoming data are saved in the upper half of 
the external SRAM and sent to the ``Zero Compression'' unit only at the 
end of data reading.
\item The ``Pad-Row Builder'' unit sorts the data according to their padrow number.
\item The ``Zero Compression'' module compresses the incoming data and writes 
them into the output FIFO.
\item The ``Command reply, Event Builder and Serialiser'' unit continuously 
takes the data from the output FIFO and converts them into serial data. 
Additionally, this module sends status information. 
Also this module is responsible for the uploading of data blocks, such as 
pedestals from the external SRAM or the firmware file from configuration EEPROM 
toward the ConcentratorBox, whenever verification reading is requested. 
In order to distinguish normal data bytes and status data bytes, the module 
frames the 8$\,\mathrm{bit}$ wide data stream to 
11$\,\mathrm{bit}$s, the leading 2$\,\mathrm{bit}$ encoding the type of 
the data byte, whereas the trailing 1$\,\mathrm{bit}$ being an end-of-frame. 
In order to reach the 50$\,\mathrm{Mbit/sec}$ effective transfer rate for 
the data bits, the LVDS lines were driven at a $11/8$ times larger frequency, 
i.e.\ at 68.75$\,\mathrm{MHz}$ because of the overhead of the frame bits.
\item The ``Command and Deserialiser'' unit parallelizes 
the incoming serial data. It decodes the received commands from the ConcentratorBox 
and decides what to send as reply: status information, the firmware file or a 
pedestal table data block.
\end{itemize}

\begin{figure}[!ht]
\begin{center}
\includegraphics[width=7.7cm]{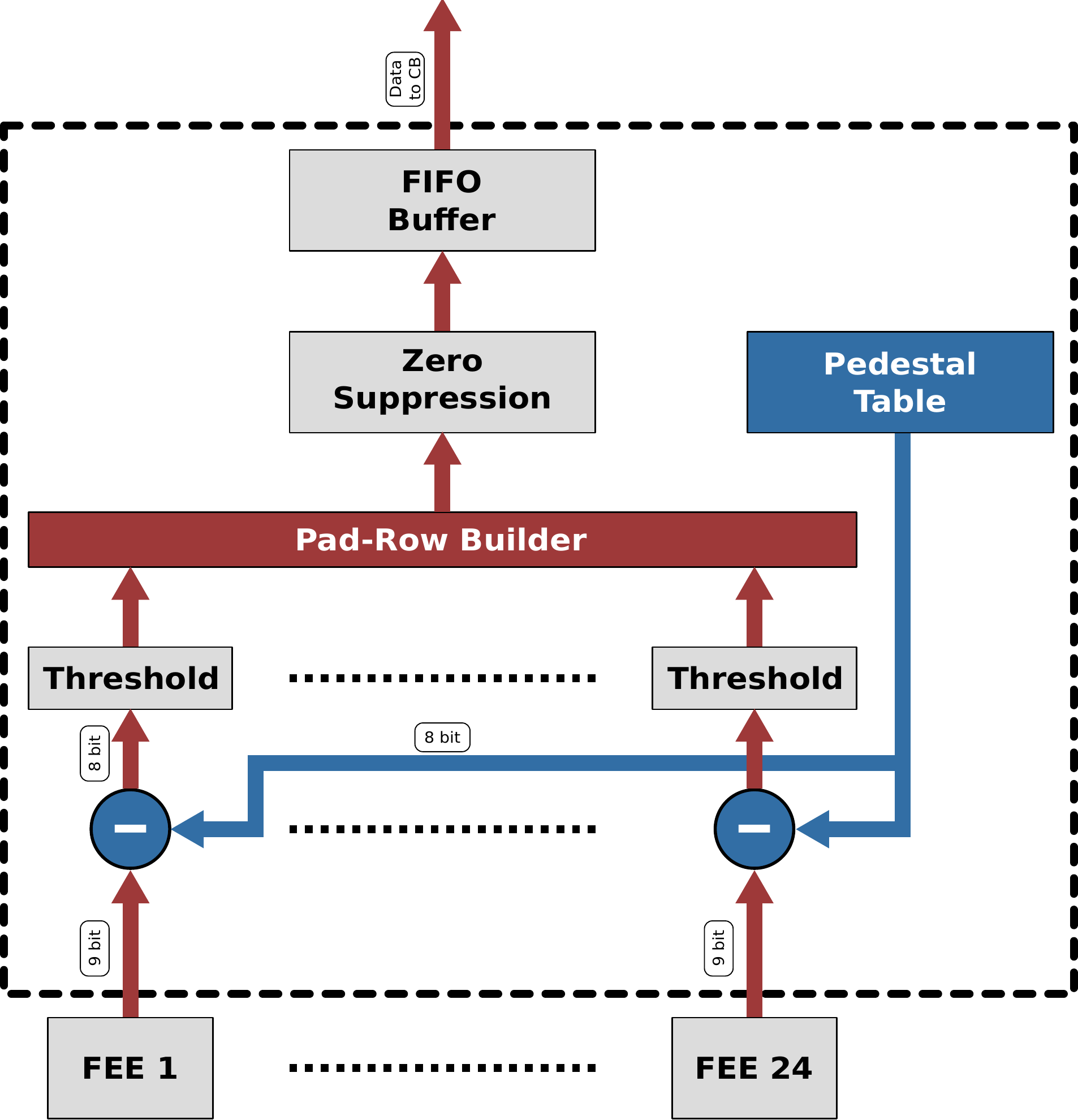}
\caption{Schematics of the data flow inside a MotherBoard.}
\label{MotherboardDataFlow}
\end{center}
\end{figure}

\begin{figure}[!ht]
\begin{center}
\includegraphics[width=7.7cm]{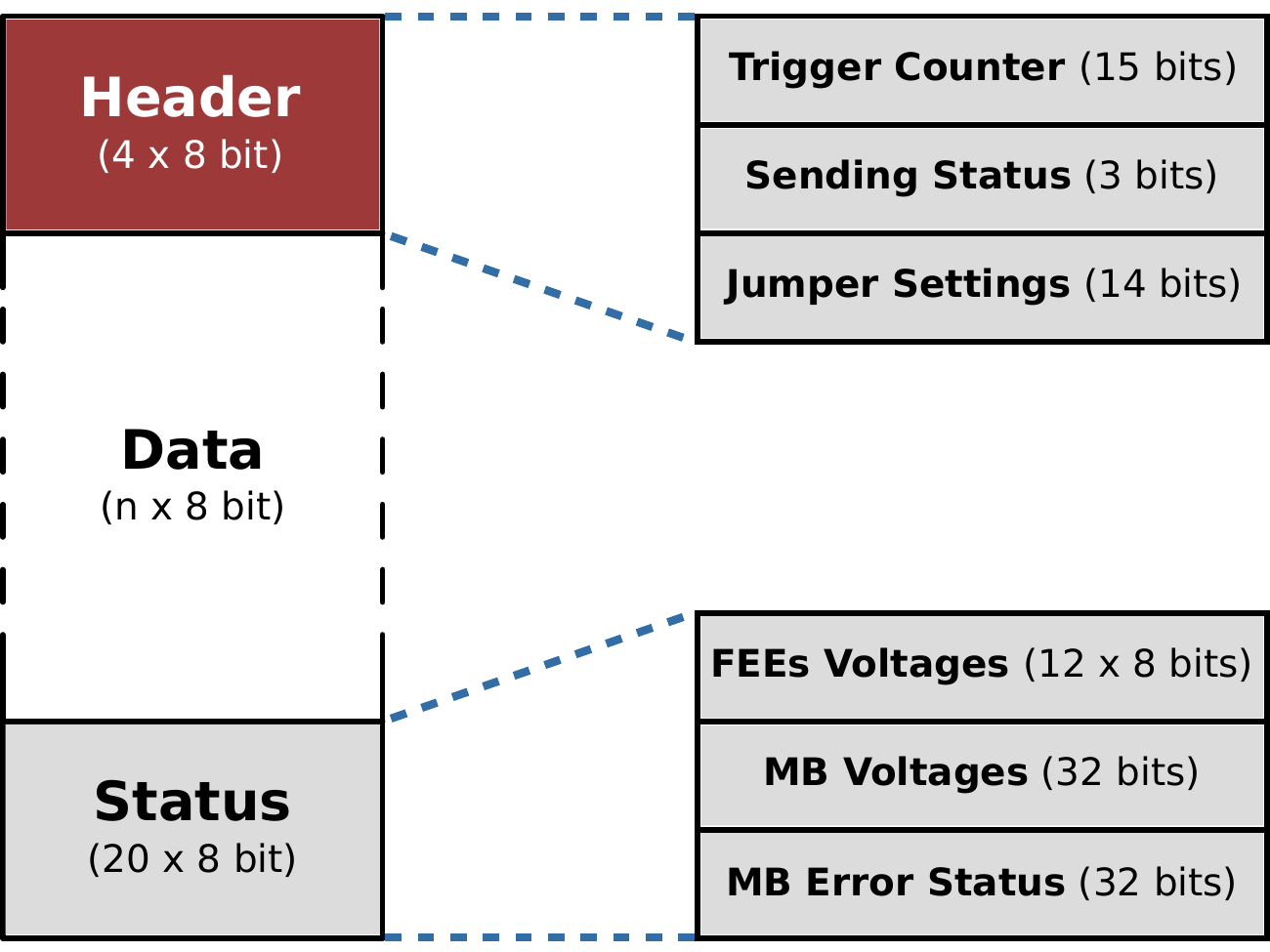}
\caption{Schematic description of the data structure on a MotherBoard output.}
\label{MotherboardDataStructure}
\end{center}
\end{figure}

The data stream of a MotherBoard is produced as follows. When the MotherBoards 
reads the 9$\,\mathrm{bit}$ ADC charge of a timeslice of a pad, the previously 
stored corresponding pedestal value is subtracted, and this data word is 
replaced by zero value in case it is below the predefined noise threshold. 
After this, the incoming 9$\,\mathrm{bit}$ data are truncated to 8$\,\mathrm{bit}$: 
if an ADC value is larger than 255, the value 255 is returned instead. 
After the above processing the data are transferred to a FIFO for each FEE card. From this 
FIFO these are read out in padrow order and sent to the zero compressor unit. 
After performing the zero compression the resulting 8$\,\mathrm{bit}$ wide 
data stream is written to the output FIFO as shown in 
Fig.~\ref{MotherboardDataFlow}. A 1$\,\mathrm{word}$ (4$\,\mathrm{Byte}$) 
event header containing the status of the jumpers and the trigger counter 
precedes the data stream. At the end of the data stream the MotherBoard can 
send a 4$\,\mathrm{word}$ (16$\,\mathrm{Byte}$) event trailer 
containing the FEE and MotherBoard voltage values, plus an error and status 
word to the ConcentratorBox (see Fig.~\ref{MotherboardDataStructure}). 
These five status words are optional, but are used by default. 
When, during data taking, the ConcentratorBox for any reason 
cannot accept data, the ConcentratorBox sends an XOFF control word to the 
MotherBoard. Receiving XOFF the MotherBoard stops sending data to the 
Concentrator. The MotherBoard temporarily stores the data in its output 
FIFO till the arriving of an XON control word. 
If during this time the output FIFO becomes full, the MotherBoard can
follow one of the two, configurable policies:
\begin{enumerate}
\item The MotherBoard continues to read out the FEE cards without loading 
data into the output FIFO. In this case the event is truncated, which fact 
is signaled in the error and status word at the end of the event.
\item The MotherBoard, after noise suppression and padrow ordering, sends 
the data temporary into the higher part of external SRAM memory. Only 
the higher part can be used for that purpose as the lower part contains 
the pedestal table. At the moment when free space becomes available in 
the output FIFO the data are transferred to the zero compressing module. 
This event of temporary storage is also logged in a warning bit of the 
error and status word.
\end{enumerate}
Normally, only the latter mode is used, the former mode is applied only in case 
of very rare calibration runs in the 512 timebin mode. 
The advantage of the second policy is that the event is transferred as a whole. 
The above sequence of reading out - pedestal subtraction - noise suppression 
- padrow ordering - zero compression - sending to ConcentratorBox is repeated until receiving 
a ``Stop Data Taking'' command from the ConcentratorBox, which signals the 
end of a data taking run for the MotherBoard.

\begin{figure}[!ht]
\begin{center}
\includegraphics[width=7.7cm]{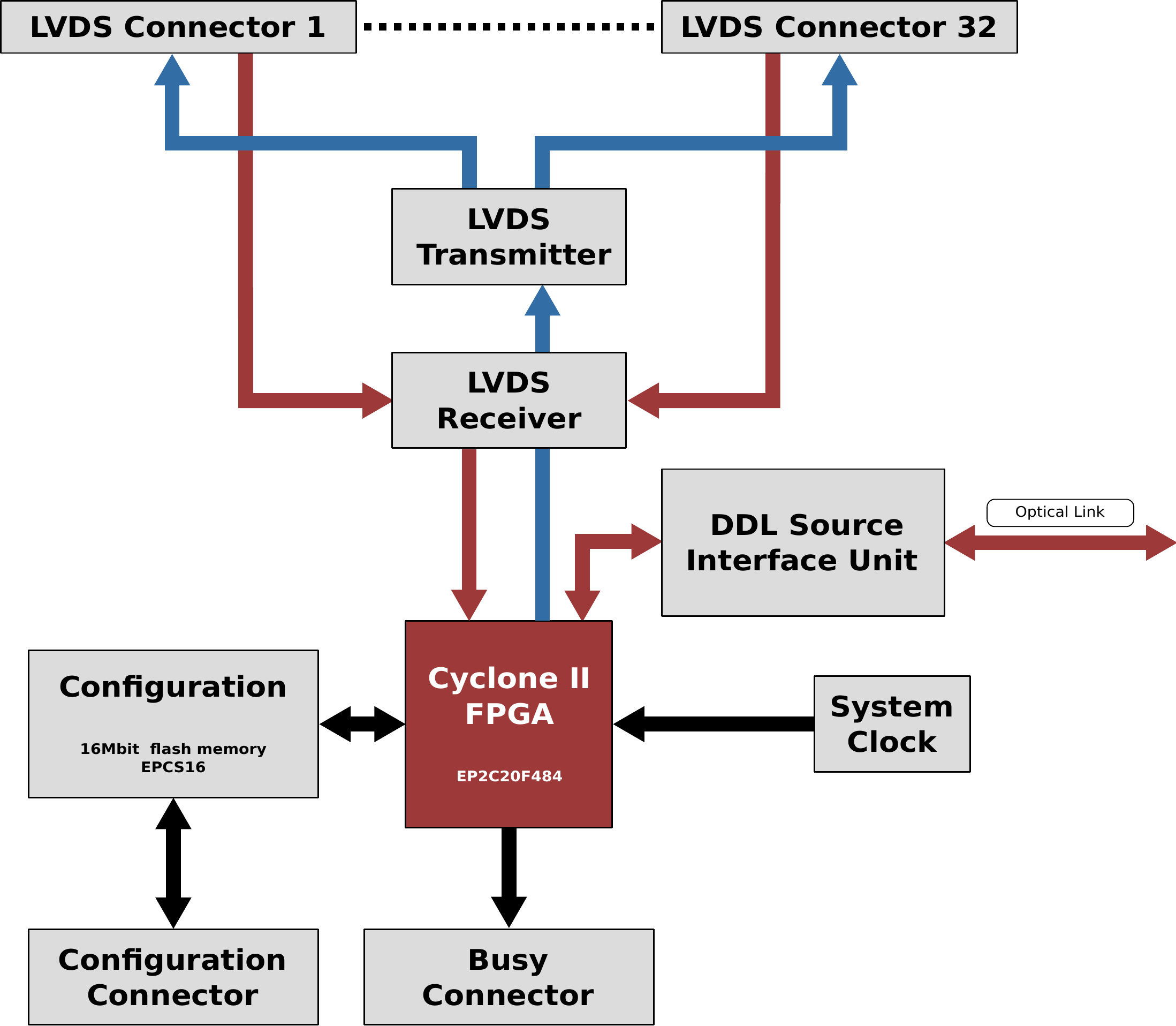}
\caption{Block diagram of a ConcentratorBox.}
\label{concentratorbox}
\end{center}
\end{figure}

\begin{figure}[!ht]
\begin{center}
\includegraphics[width=7.7cm]{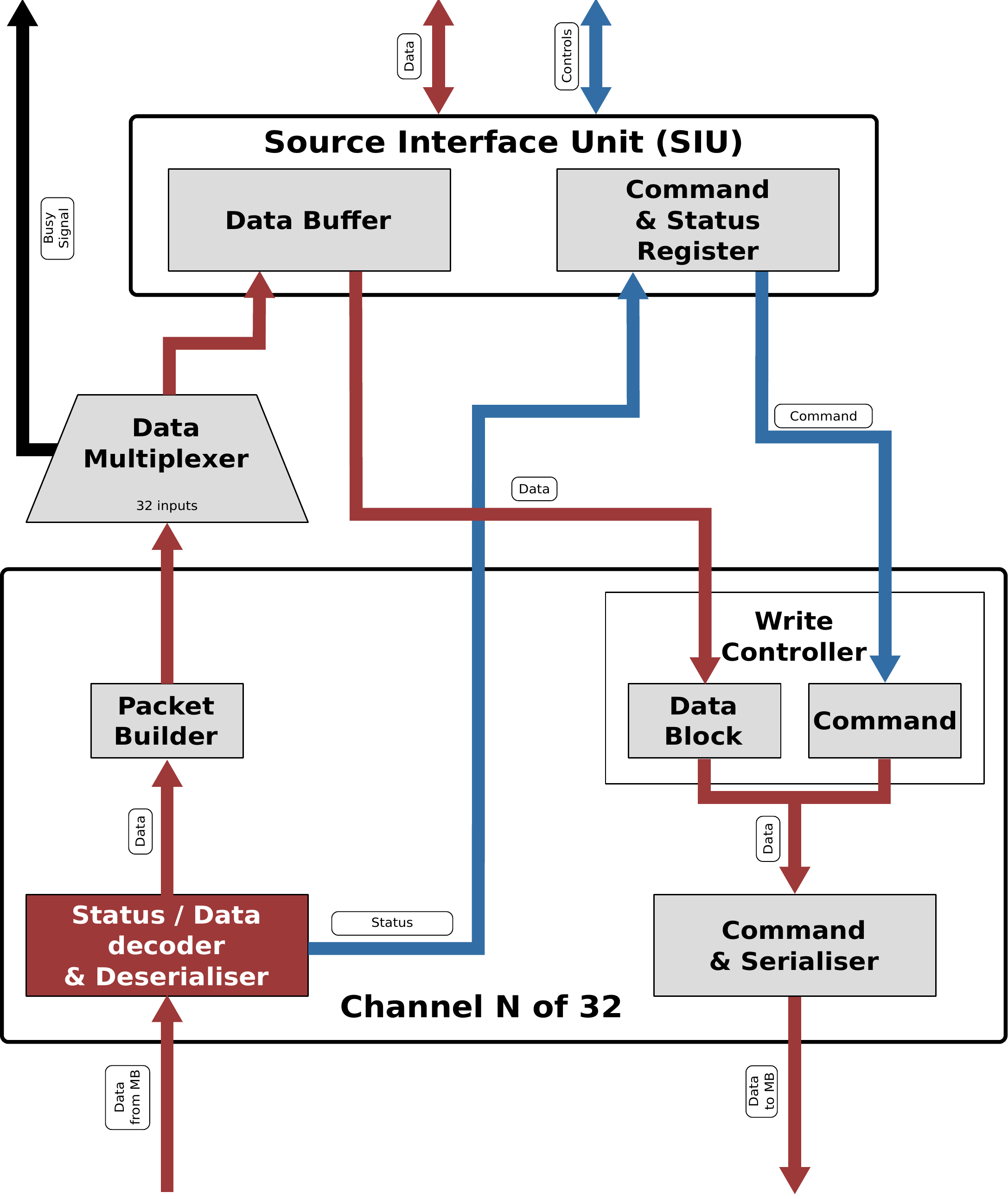}
\caption{Block diagram of ConcentratorBox firmware.}
\label{concentratorboxfw}
\end{center}
\end{figure}

\begin{figure}[!ht]
\begin{center}
\includegraphics[width=7.7cm]{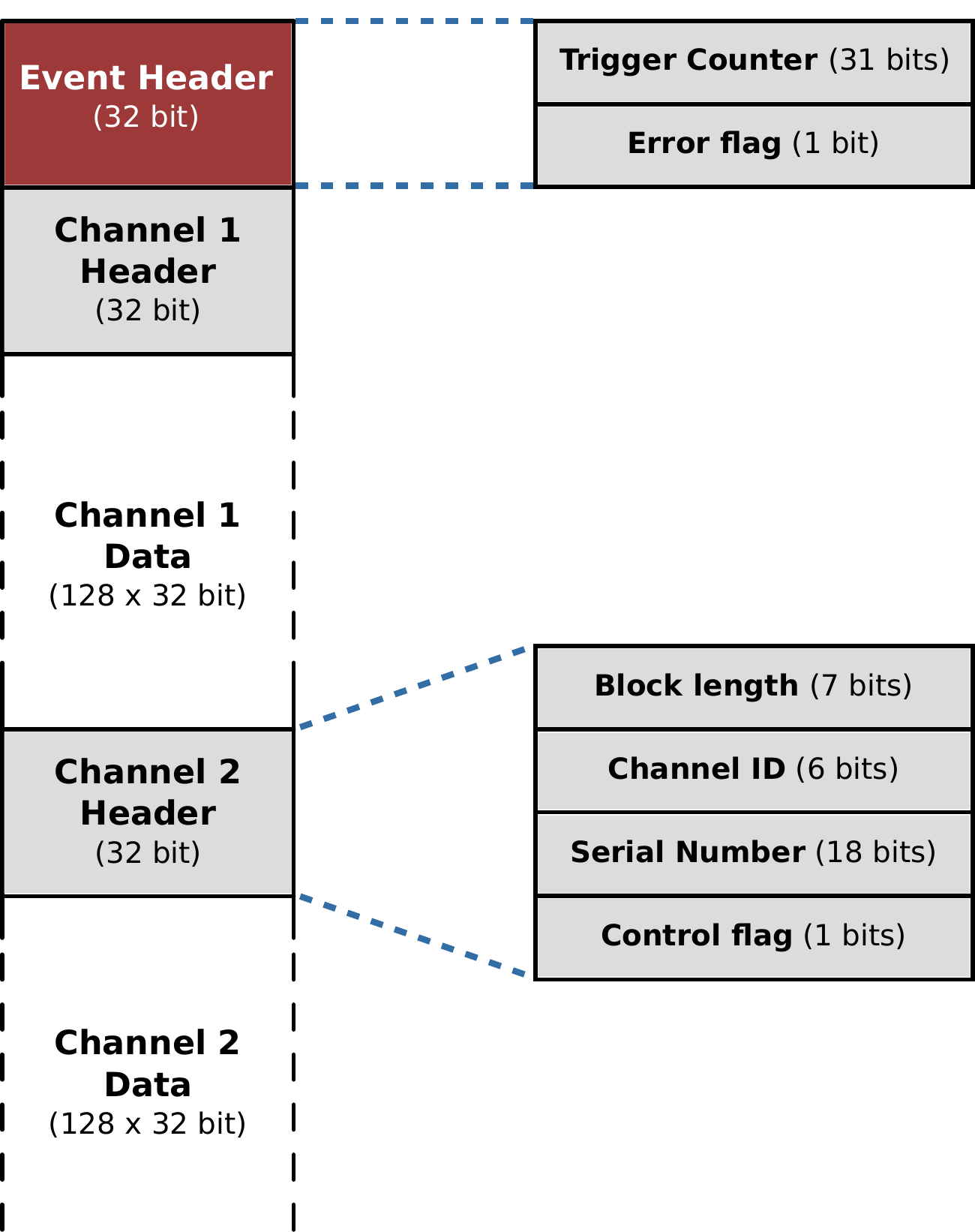}
\caption{Data structure at the output of a ConcentratorBox.}
\label{ConcentratorDataStructure}
\end{center}
\end{figure}

The ConcentratorBox receives data packets from the MotherBoards via 32 LVDS 
channels and forwards their serialized data onto a DDL line, as shown in 
Fig.~\ref{concentratorbox}. 
The firmware of a ConcentratorBox (Fig.~\ref{concentratorboxfw}) has two main parts: 
the serializer/deserializer for each channel and the Data Formatter, 
which creates data structure to be fed to the DDL. For each channel 
one receiver and one transmitter logic is implemented. 
The receiver logic recognizes the start and end of event frame, performs 
serial to parallel data conversion and buffers the data in the receiver FIFO.
The transmitter logic reads data from the transmitter FIFO, performs 
parallel to serial conversion, and attaches start of frame and end of 
frame symbols. Both logics ensure the flow control between the MotherBoard 
and the ConcentratorBox. The input packets are buffered in FIFOs. The readout 
of the FIFOs is organized in cycles. 
In each cycle the firmware reads maximum 128 pieces of 4$\,\mathrm{Byte}$ 
data words from each channel's receiver FIFO in a sequential way. If the 
FIFO of a channel is empty then the given board is skipped. A so-called 
mini-header is attached to each data packet containing the identification 
number of the source MotherBoard and the length of the block. These packets 
are called ``LVDS trains''. The readout cycle 
continues till the end-of-frame signal is reached in all channels. Before 
sending the first mini-header an event header is sent to the DAQ. The structure 
of the data stream transferred to the DDL link is showed in 
Fig.~\ref{ConcentratorDataStructure}. Upon the arrival of the first data to the ConcentratorBox, 
a TTL output line is set to logical one level (Busy signal), and this signal 
persists until the ConcentratorBox recognizes that all data from all of its 
MotherBoards are drained out. The Busy signal is used for trigger level 
synchronization of the subdetectors.

The DDL transfers data from ConcentratorBox to RORC plugged in one of PCI-X or PCI-Express slots of the DAQ 
computer. The DDL works using a push-data mode protocol, i.e.\ after starting data 
collection the ConcentratorBox sends the data to the RORC when they are produced, 
without waiting for any request. The RORC's task is to transfer the data into 
DAQ PC's memory via the PCI bus. To fulfill this task the RORC has a FIFO up to 128 
entries which can be filled with PCI addresses and lengths of memory blocks 
where the data blocks can be written to. When a data block arrives the RORC 
takes the next address in the FIFO and writes the block there using 
DMA (Direct Memory Access), without processor intervention. For the DMA, it is 
necessary to provide continuous physical memory pages. To assure this, we use a 
special memory tool: the Physmem package, which gives access to a fixed part 
of the physical memory of a Linux machine for unpaged, un-swapped physical 
I/O. This memory is outside of the memory assigned for the operating system. 
However it can be reached by both the kernel and the user processes via the 
Physmem driver. The package's driver is loaded as a kernel module at boot time, 
and care must be taken to limit the memory used by the kernel using boot 
arguments in order to allow space for the Physmem above Linux memory. 
The driver provides the user with the physical address and size of the 
continuous memory which can be used as the destination area of read out data.

During one DMA, only one data page can be written. Data pages that belong 
to the same event are transferred over the DDL in one or more DDL blocks. 
Each block can be up to $4\times(2^{19}-1) = 2097148$ bytes. When a data 
block is transferred into the Physmem memory the RORC puts two control words 
to a predefined area of memory where it specifies the length of the block 
and signals if the block is the last one of an event.

For RORC functionality the RORC's driver module has to be loaded. 
This driver produces the mapped user addresses of the RORC registers. 
The pertinent addresses are 
used for accessing the RORC registers and address FIFO. Using these registers the 
DAQ program initializes the ConcentratorBoxes and MotherBoards, can send 
pedestal data, while using the address FIFO it can follow the course of the 
data acquisition.

Presently, the TPC Data AcQuisition is steered by the Central DAQ software, 
and is ran on a single Central DAQ PC with the following parameters: 
X7DB8-X motherboard, 64$\,\mathrm{bit}$, 
total 8 cores of Intel Xeon CPU at 2$\,\mathrm{GHz}$, 8$\,\mathrm{GByte}$ 
memory, 6 PCI-X card slots, 4 USB ports, 1 serial port. The computer runs with 
a 64$\,\mathrm{bit}$ SLC6 Linux operating system. The memory available for the Linux kernel is restricted 
to 4$\,\mathrm{GByte}$ using a kernel argument. The remaining, upper 4$\,\mathrm{GByte}$ 
memory is given to the Physmem kernel module, for receiving the subevent data. 
The built event data are written to a fast temporary disk buffer, consisting 
of a RAID0 array of 5 pieces of 2$\,\mathrm{TByte}$ hard disks, providing about 
5 times larger bandwidth for writing than a single hard disk.

The data taking cycle of the full TPC system causes approximately 12$\,\mathrm{ms}$ deadtime, 
determined by the present setting of TPC FEE cards.

\section{Further subdetectors}
\label{psdcamactofelectronics}

A number of further subdetectors (PSD, beamline detectors, ToF detector) also 
needs to be read out along with the tracking data of the TPC system described 
in Sec.~\ref{tpcelectronics}. These, however, only produce small volume 
data, merely 230$\,\mathrm{kByte}$ without any compression. Also, their deadtime is 
smaller than that of the TPC system. The organization of the readout of these 
subsystems are summarized in the following subsections.

\subsection{The readout electronics of the PSD calorimeter}

The PSD calorimeter \cite{psd} is a very important component of the NA61 
experiment for the nucleus-nucleus collision data taking periods. This 
detector is designed for the determination of collision centrality by measuring 
the energy of the spectator nucleons, i.e.\ of beam nucleons not participating 
in the collision. Its design parameters were set such that its energy 
resolution allows centrality determination with a precision of $\pm$1 spectator 
nucleon at the lowest used beam energy. The calorimeter 
consists of 44 modules covering a 120$\times$120$\,\mathrm{cm}^{2}$ transverse area, 
and the longitudinal depth covering 5.7 nuclear interaction lengths. The 
modules of the calorimeter consist of a sequence of lead and plastic 
scintillator plates, which are grouped into 10 longitudinal sections for 
shower profile monitoring, the light yield of each section being read out 
by an MAPD (Micro-pixel Avalanche PhotoDiode). The MAPD signals are stored in time samples and are read out 
by PSD MotherBoards. These were designed to have similar interface towards the 
Central DAQ as the TPC MotherBoards, connected to the system via the 
ConcentratorBoxes. The raw data volume of the PSD subevent is 200$\,\mathrm{kByte}$, 
and its detector readout has a deadtime of 2.5$\,\mathrm{ms}$.

\subsection{The readout electronics of the beamline detectors}

The beamline detectors, such as the beam scintillators, \v{C}erenkov 
counters and beam position detectors are read out using legacy commercial 
CAMAC ADC, TDC and scaler modules, altogether $\sim$300 channels. 
A CAMAC-to-VME bridge (CES VIC8251) 
provides connection to a dedicated VME crate containing a VME controller 
(CES FIC8234), which runs simple low level readout software. 
Since the FIC8234 can receive an external trigger interrupt upon the 
detection of a physical collision, this feature is used for notification 
of the low level readout software to move the event data via the CAMAC-to-VME 
bridge into a VME memory module (MicroMemoryInc MM6390), in which these 
subevents are stored in a ring buffer. 
This low level DAQ system is self-contained and can run independently 
from Central DAQ. The Central DAQ PC may access the pertinent MM6390 VME 
memory module via a VME-to-USB bridge (CAEN V1718) in order to poll 
for new subevents. Whenever a new subevent arrives, the Central DAQ PC copies 
it into its own ring buffer. 
A data flow blocking mechanism is implemented via a VME commandable register 
unit (CES RCB8047 CORBO). The Central DAQ can use it to pause the event flow 
recorded by the low level system in the rare cases of transient bandwidth 
problems with the moving of the subevent data. 
The data volume of this subevent is 
merely 6.5$\,\mathrm{kByte}$. The deadtime of this subdetector readout is 
$11\,\mathrm{ms}$.

\subsection{The readout electronics of the ToF system}

A quite similar setup is used to read out the 2080 TDC (Time-to-Digital Converter) and 
QDC (Charge Analog-to-Digital Converter) channels of the ToF system. 
The legacy Fastbus based TDC and QDC units are linked using a Fastbus-to-VSB 
bridge (CES LDA9212) to a dedicated VME/VSB crate steered by a VME/VSB 
controller (CES FIC8234), running a low level readout software. 
The Fastbus TDC and QDC units 
are pre-triggered using the 30$\,\mathrm{ps}$ time resolution start scintillator signal, S1\_1, 
of the experiment, and the pre-trigger may or may not be confirmed by a 8$\,\mathrm{ns}$ 
time jitter Main Trigger signal generated by the trigger logic using 
further beam counters within a timeout of 300$\,\mathrm{ns}$. If the Main Trigger was not present, 
the TDC and QDC data are cleared using a Fast Clear signal causing only a 
1$\,\mathrm{\mu sec}$ deadtime. Whenever the Main Trigger signal 
confirmed the pre-trigger, the low level readout software on the FIC8234 
controller is notified via its external interrupt port. The TDC and QDC data are moved 
via the Fastbus-to-VSB bridge into the internal memory of FIC8234, 
and these subevents are stored in a ring buffer scheme. The Central DAQ PC 
polls this ring buffer for new events and moves them away via a 
VME-to-USB bridge (CAEN V1718) and propagates them to its own ring buffer in memory. 
The data flow control to the Central DAQ is again realized via a VME configurable 
register unit (CES RCB8047 CORBO) in order to prevent the low level ring 
buffer from overwriting itself before the moving of the subevents by the 
Central DAQ takes place. The data volume of this subdetector is 21$\,\mathrm{kByte}$, 
and its deadtime is 8$\,\mathrm{ms}$.

\begin{figure}[!ht]
\begin{center}
\includegraphics[width=7.7cm]{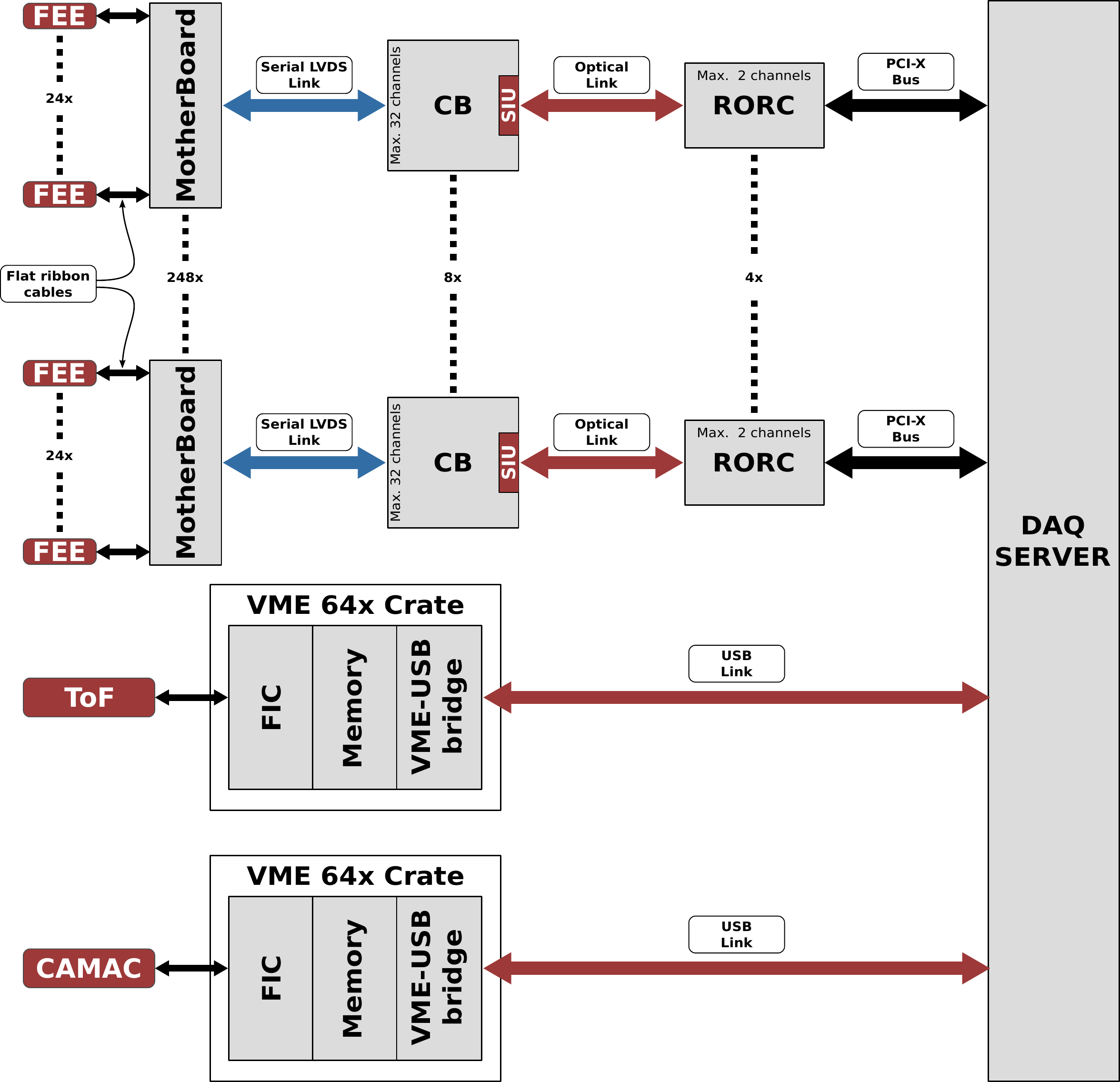}
\caption{Overview of the entire data acquisition system of NA61 along 
with the new TPC readout system, PSD readout system and the legacy beam detectors and ToF system.}
\label{DAQOverview}
\end{center}
\end{figure}

An overview of the organization of the entire NA61 readout system including 
the legacy beam detector and ToF readout is outlined in Fig.~\ref{DAQOverview}. 
The legacy FEEs of the ToF detectors as well as the beamline detectors are 
foreseen to be upgraded using the new DRS technology \cite{drs} in the near future, 
and their data transmission scheme is foreseen to be adapted to that of 
the TPC system discussed in Sec.~\ref{tpcelectronics} based on 
DDL \cite{ddl} technology.

\section{Signal and data flow}
\label{dataflow}

\begin{figure*}[!ht]
\begin{center}
\includegraphics[width=16cm]{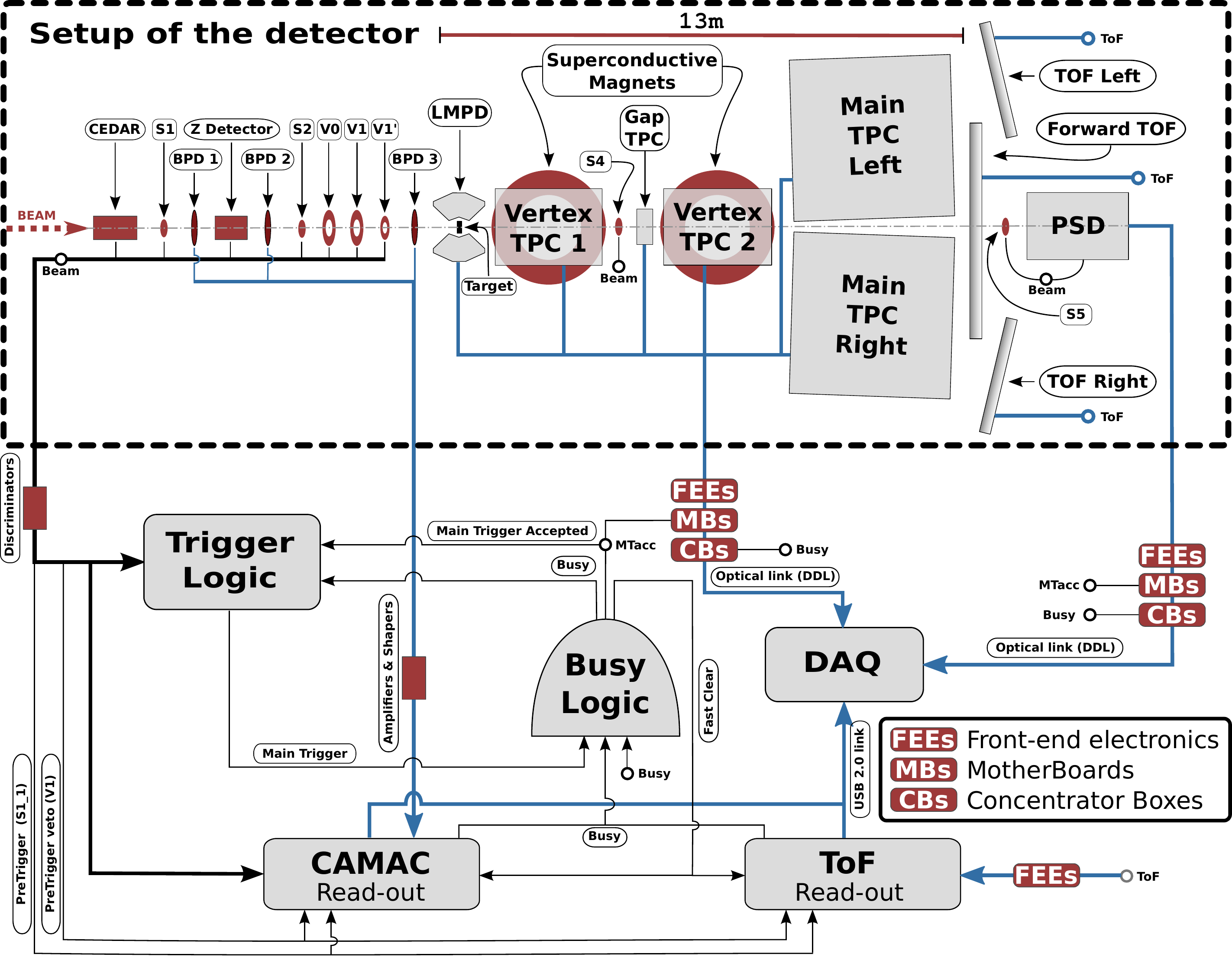}
\caption{Summary figure showing the signal and data flow 
lines within the NA61/SHINE experimental setup. The work cycle is started 
by the signal of the best time resolution photomultiplier of the S1 counter (S1\_1) 
as a pre-trigger, and may be vetoed by the signal of V1 as a pre-trigger-veto 
in order to filter out deflected beam particles. 
The rest of the beam counters are used by the Trigger Logic to build a Main Trigger signal 
confirming the pre-trigger. In case all the subsystems are in ready state, 
the Busy Logic lets the Main Trigger pass, triggering the actual data readout. 
The data flow lines from the subdetectors are also indicated.}
\label{dataflowsummary}
\end{center}
\end{figure*}

As mentioned previously, in the full experimental setup the data taking is 
controlled via a set of trigger signals from beam counters and Busy signals 
from subdetectors. These communication lines, along with the actual data flow 
lines within the experimental setup are summarized in Fig.~\ref{dataflowsummary}. 
The S1\_1 signal serves as a timing reference and a 
pre-trigger for the detector complex. The signal of the V1 veto counter, used 
for filtering out already scattered beam particles just upstream of the target, 
is applied as a pre-trigger-veto. The rest of the counter signals 
(S2, V0, V1', CEDAR, Z Detector, S4, S5) have somewhat worse time resolution 
and are used by the Trigger Logic coincidence unit to build the final Main 
Trigger signal, confirming the pre-trigger. If all the subsystems are in a 
state that is ready for data taking, the Busy Logic lets the Main Trigger pass, 
and thus generating the Main Trigger Accepted signal in order to start 
actual reading out of the detector data. The pathways of the data flow from 
the subdetectors, such as the TPC, the PSD, the ToF and beam counter systems 
are also indicated.

\section{Data synchronization}
\label{synchronization}

The synchronization of the deadtime of the subdetectors for each event is 
done by a Busy Logic system. This vetoes the trigger signal distributed to all the 
subsystems in case any of the subdetectors are not in their ready states, 
which is determined via a Busy signal from each subdetector. In addition, 
the trigger is also blocked if any of the fast systems (beam detectors, or 
ToF system) received a pre-trigger, i.e.\ an S1\_1 signal not vetoed by V1.

As the NA61 DAQ system is organized in a push-data mode, i.e.\ the low level 
systems receive the data taking trigger independently and they push their 
data asynchronously to a higher level, some kind of protection 
mechanism is needed for checking against any mixing of subevents in the parallel 
data streams. Initial synchronization of the readout systems of all 
the subdetectors occurs at the start of a data taking run, when all 
the subdetector trigger counters are reset. In order to maintain this 
synchronization, we use the fact that the SPS beam is provided 
to NA61 experiment in $\sim$10$\,\mathrm{sec}$ long ``spills'' with at least 
5$\,\mathrm{sec}$ break in between them (in the most usual data taking conditions 
the duty cycle is about 30\%). 
During the spill breaks, all the data channels toward the Central DAQ are emptied. 
Upon detecting a stop of event 
flow for 3$\,\mathrm{second}$s, the trigger counters in the subevent data channels 
are cross-checked as they should be equal if no false trigger has been received 
by any subdetector and if no subdetector failed to receive a true trigger. In 
case the trigger counters are equal, the corresponding identifier of the 
last recorded event is marked as the last good one, and is approved for 
archiving. In the rare cases when this check does not pass, the data taking run 
is halted, the event stream is deleted back to the last good event, the Central 
DAQ issues an error message of ``trigger skew'', and a completely new data 
taking run is started with all the subdetectors fully reset. Such a recovery 
run stop-and-start takes about 1 minute.

A further crash scenario can occur when the finalization of the subevent by one 
of the subsystems is unfinished, i.e.\ the end-of-subevent trailer in a given 
channel does not arrive. In such case, the Central DAQ issues an 
``event reading timeout'' within 1$\,\mathrm{second}$ and a completely 
new run is started with subdetectors reset.

A very rare but very harmful crash scenario is when a subsystem starts data taking 
e.g.\ due to a false trigger, but is not able to send the subevent header 
to the Central DAQ due to some internal inconsistency. In such case the 
Central DAQ cannot know that the recording of a new event has been started 
and therefore cannot time out. However, 
since the data taking cycle has begun, the given subsystem sets its busy 
signal thus blocking the arrival of any further trigger signals. In such 
scenario the data taking would hang completely without any particular 
error message until human intervention could detect and resolve the problem. Therefore, a custom made TTL 
register module was built in order to monitor the busy signals of all the 
subdetectors. The pertinent register can read 16 TTL signals which are 
visible to the Central DAQ PC via its RS232 serial port. Whenever the 
subevent flow comes to a standstill the busy register is checked that 
none of the busy signals remained in the ``true'' state. In case any of the busy 
signals remain in ``true'' 1$\,\mathrm{second}$ after the halt of the event flow, a 
``a busy signal got stuck'' error is issued by the Central DAQ and a completely 
new run is started with subdetectors reset. In order to maintain galvanic 
isolation of the Central DAQ PC, an optical RS232 coupler is used.

The occurance rate of any kind of data inconsistency condition in the data 
stream during stable data taking is seen to be about 0 to about 5 times 
a day, which is considered to be well tolerable. These failures are normally 
traced back to some infrastructural problem such as power supply or 
cooling water service instability.

\section{The readout/online software}
\label{onlinesoftware}

\begin{figure}[!ht]
\begin{center}
\includegraphics[width=7.7cm]{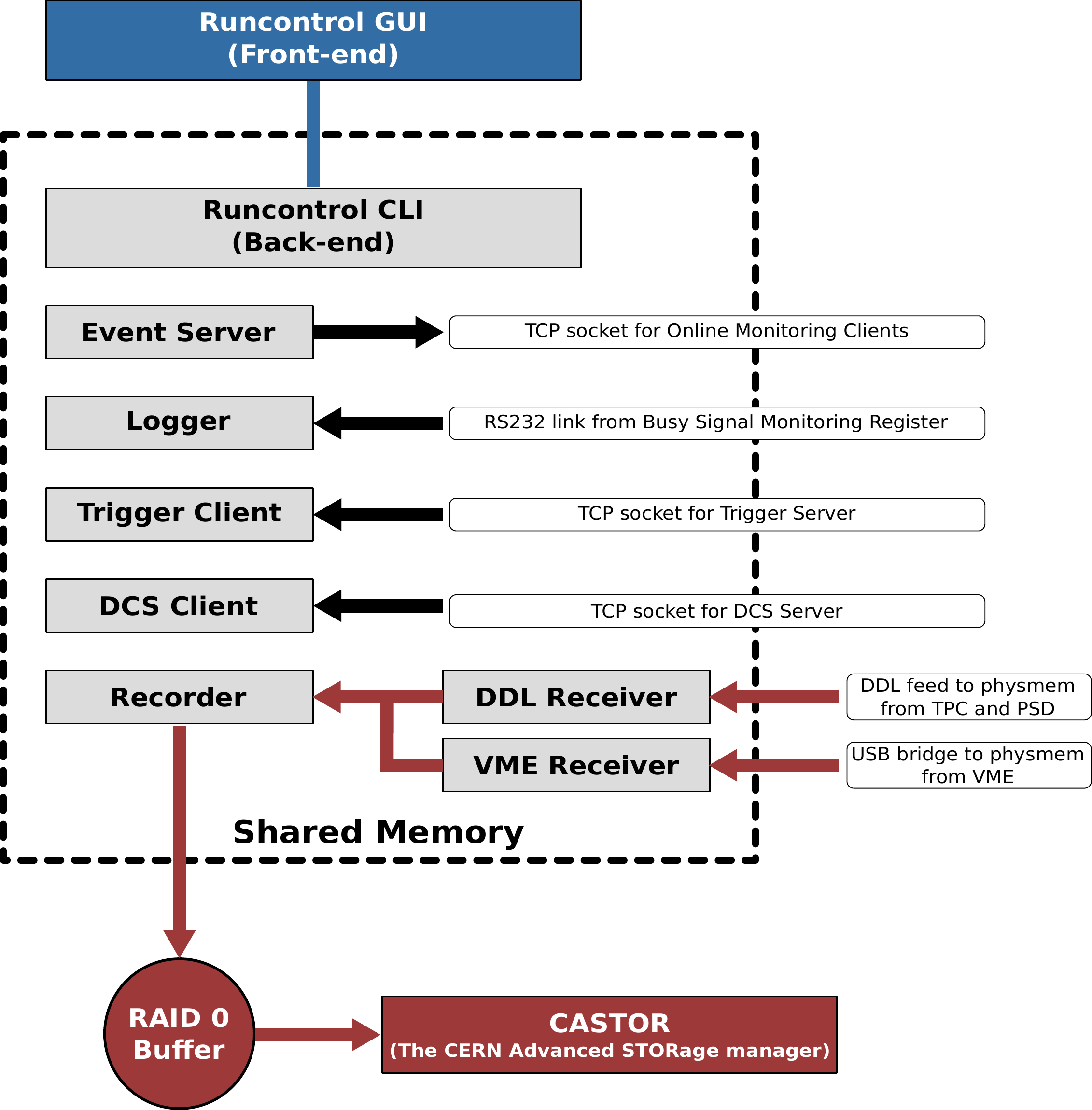}
\caption{Sketch of the data flow of the Central DAQ software, 
the Runcontrol, during data taking run. The figure also depicts the 
process hierarchy, consisting of 9 individual subprocesses. 
The DDL Receiver and VME Receiver lay the arrived data into Physmem pages, 
and these page addresses are passed to Recorder for event building. 
After the Recorder has built the events onto disk, it the recycles 
addresses of the free Physmem pages to DDL Receiver and VME Receiver.}
\label{swsketch}
\end{center}
\end{figure}

The structure of the Central DAQ software, called Runcontrol consists of the 
following 9 subprocesses.
\begin{enumerate}
\item \emph{GUI}: the GUI of the software is a Tcl/Tk \cite{tcltk} wrapper script, providing a 
user friendly interface.
\item \emph{Runcontrol}: the core software under the GUI is written in C language, running 
autonomously. Its user interface is a simple shell-like command line 
interpreter based on GNU Readline Library \cite{readline}. For experts, it is possible 
to use this shell-like command line environment for debugging mode, however, normally 
it is handled by the GUI, emulating user commands which are sent and received 
between the command line interpreter and the GUI via unix pipes and log files.
\item \emph{Event Server}: whenever the core of Runcontrol is started, an event server process 
is created. This is a fork-server and can serve up to 16 monitoring clients 
for online event display.
\item On top of the above, the following processes are launched upon start of a data taking run:
\begin{enumerate}
\item \emph{Logger}: this is logger and consistency checking process. It is also responsible 
for checking the trigger counter synchronization of all subsystems whenever 
there is a pause in the data stream. In case the trigger counters on the 
incoming DDL and VME channels are inconsistent, a ``trigger skew'' error is 
issued and a completely new run is started with subdetectors reset. Also, the busy signals 
of the subsystems are monitored by this process using our custom made register 
unit accessible via the serial port.
\item \emph{Trigger Client}: this is a process for communication with the 
Trigger System. This client retrieves scaler counts and other trigger 
summary information at the end of each spill, and writes it into the data stream 
for monitoring purposes.
\item \emph{DCS Client}: this is a process for communication with the DCS 
(Detector Control System), responsible for monitoring the slowly changing 
detector parameters, 
such as temperatures, high voltages etc. This process periodically retrieves 
logs from the DCS system and appends to the event data stream for 
monitoring purposes.
\item \emph{Recorder}: this process is responsible for collecting the subevents 
on a first come first served basis from the data channels of the receiver 
processes. It collects the subevent data addresses in an event data pool for each 
incoming subdetector data channel. If the end-of-event trailer arrives in each data channel, 
it builds the event directly onto disk following an internal consistency 
check. Upon request, it also duplicates in memory the event being built for the 
Event Server process. The Recorder process has two further processes subordinate to 
it for efficient parallelization of the data receiving procedure.
\begin{enumerate}
\item \emph{DDL Receiver}: this process waits for subevent data delivered by 
the RORC cards and forwards their address on a first come first served basis to 
the Recorder process. The RORC cards deliver the event data to the 
pre-allocated Physmem area without CPU intervention. The addresses of the 
available Physmem pages, returned by the Recorder, are filled by this process 
to the RORC cards internal FIFO. In the pathological case when the Recorder cannot 
build the events fast enough, the RORC cards may run out of free Physmem 
page addresses. In that case, the RORC cards stall the data stream via 
issuing an XOFF signal towards the DDL lines. The data stream is restarted 
by XON, when addresses of new free Physmem pages arrived into the internal 
FIFO of the given RORC card.
\item \emph{VME Receiver}: this process polls for subevent data in the VME 
memory of the low level systems, accessed via the CAEN V1718 VME-to-USB bridge 
units. Whenever a new subevent is detected in any of the subsystems, it is 
copied to the Physmem area allocated for this process, is stored in a 
ring buffer, and the subevent address is passed to the Recorder process. 
In the pathological case when the Recorder cannot drain the subevents fast 
enough, the following event triggers are blocked by commanding our 
VME controllable TTL register units (CES RCB8047 CORBO) to issue a busy signal.
\end{enumerate}
\end{enumerate}
\end{enumerate}
The non data intensive inter-process communication of the 8 subprocesses 
of the core of the Runcontrol software is solved using standard Linux shared memory. 
For the data intensive inter-process buffering, the Physmem area is used 
instead, in order to avoid unnecessary overhead by the Linux kernel. 
The data flow of the Central DAQ software during data taking is illustrated 
in Fig.~\ref{swsketch}.

The DAQ software produces its final output data stream as a sequence of 
1$\,\mathrm{GByte}$ files, called chunks, for easier handling. Whenever the 
currently being written chunk exceeds 1$\,\mathrm{GByte}$, the corresponding 
chunk is closed and the following event is written to a new chunk. The chunks 
are written to the RAID0 buffer disk in order to reach the necessary 
bandwidth for writing (the writing bandwidth to a RAID0 disk is larger than 
to a single disk).

The DAQ software is supplemented with a set of failure-tolerant system of 
scripts, responsible for final treatment of the chunks.
\begin{enumerate}
\item Online QA (Quality Assessment) of the newly arrived raw data chunks via a dedicated 
online QA software for fast integrity checks and for acquiring histograms of important 
basic quantities for monitoring and sending them for online display. 
That piece of software is using the standard NA61 offline software framework, 
called Shine \cite{shine}. Subsequently, the properties of the produced control 
histograms are also automatically verified by a simpler program 
(``Electronic Shifter''). In this way, an alarm can issued whenever e.g.\ a 
well-defined detector failure occurs. The QA procedure is performed via two 
independent processes for efficient parallelization.
\item Copying of the certified raw data chunks to the CASTOR archiving system \cite{castor} of 
CERN. This is performed via two independent processes for efficient parallelization.
\item Verification of the raw data chunks on tape. The CASTOR system provides 
the adler32 checksum and size of the chunks migrated to tape. These two 
verification parameters are cross-checked with that of the chunk on the Central 
DAQ PC's RAID0 buffer disk by a script. The local copy on the Central DAQ PC 
is deleted whenever the verification passes. In the rare cases when verification 
fails, the raw data chunk is kept on the Central DAQ PC for human intervention.
\end{enumerate}

\section{Raw data structure}
\label{rawdatastructure}

Book-keeping of the DDL page structure along with the ``LVDS train'' substructure 
of the data stream is performed by the DDL Receiver process. 
These are unfolded from the data stream at the 
event building level upon writing to disk, in order to have contiguous binary 
data from each MotherBoard, free from all the data frames used only for 
communication protocols between MotherBoard and ConcentratorBox, or 
ConcentratorBox and Central DAQ. The subevent data from the beam detector 
system or the ToF system are already built by the low level FIC controllers, 
and are simply received by the VME Receiver process and passed over for 
event building without modification.

The raw data structure written by the Central DAQ is organized as follows. 
Each event is written on disk after each other as a Record, consisting of a sequence of 
4$\,\mathrm{Byte}$ integer words, the first word encoding the number of 
words in the Record (inclusive), followed by the data words of the Record. 
The data words are organized in BOS (Bank Objective System) banks, consisting of a 4$\,\mathrm{word}$ 
header and a sequence of data words within the BOS Bank. The first BOS header 
word is the identifier of the type of subdetector data, the second BOS header 
word is a geographical detector identifier, the third BOS header word 
is reserved (set to zero), and the fourth BOS header word encodes the 
number of following data words (exclusive) within the BOS Bank. The 
4$\,\mathrm{Byte}$ words in the event data stream, whether interpreted as 
integers or floating point numbers, are always encoded in the ``network byte order'', 
also called ``Big-Endian format''.

The most important BOS Banks are the followings: 
EVEH (event header data),
CAMC (beam detector data),
TOFD (ToF data),
MBRD (TPC MotherBoard data), 
MBLV (TPC MotherBoard housekeeping parameters, i.e.\ low voltages, temperatures), 
PSDD (the same as MBRD for the PSD detector),
PSLV (the same as MBLV for PSD detector).

The used raw data structure based on Record and BOS Banks proved to be 
a flexible way for extension of the detector system or for deprecation of 
old subdetectors.

The size of a ``black event'' (event without pedestal subtraction, noise 
suppression and zero compression) is 100$\,\mathrm{MByte}$ if the TPC drift 
time is sampled in 512 timebins, whereas 50$\,\mathrm{MByte}$ in case of the 
typically used 256 timebins. For a normal event (event with pedestal subtraction, 
noise suppression and zero compression), the typical size is 1.5-5$\,\mathrm{MByte}$, 
slightly depending on track the multiplicity in the event.

\section{Performance}
\label{performance}

The presented readout and Data Acquisition system was set first in operation 
in 2009, with the start of the data taking program of NA61/SHINE. After 
initial debugging, it served as a very reliable system throughout the extended 
data taking periods of 2-5 months per year, and so far only minor upgrades 
were needed. The total deadtime, 12$\,\mathrm{ms}$, of 
the system proved to be sufficient to complete the experimental 
program in a timely manner. Improvement on the deadtime would not provide 
further significant benefits as presently our data taking rate is limited by the 
available beam intensity for most of the collision types as well as 
the limited target thickness in order to suppress secondary collisions. 
Therefore, the construction is considered quite mature and stable. 
The acquisition system was used to record a large amount of physics 
data on 29 kinds of reactions, with the number of events ranging from 1 
to 40$\,\mathrm{M}$, totaling up to 194$\,\mathrm{M}$ recorded physics 
events during the years 2009-2015. It is expected that a further 20 kinds of 
reactions will be recorded through 2015-2019, typically comprised of 4$\,\mathrm{M}$ 
events each.

\section{Concluding remarks}
\label{concludingremarks}

The design concepts and the used technologies of the readout system of 
the NA61/SHINE experiment at the CERN SPS were summarized. Special emphasis 
was given to the readout of the large, 40$\,\mathrm{m}^3$ TPC system; 
the most important detector component in the complex. 
The readout technology is largely 
based on the DDL data transmission, also used in ALICE. The present system 
provides possibility for reading out physics events at a rate of 83$\,\mathrm{Hz}$, 
which satisfies the present needs given the upper safety limit of usable 
beam intensity. The raw event size is 50$\,\mathrm{MByte}$s, which after compression 
reduces to 1.5-5$\,\mathrm{MByte}$ depending on the number of particles 
in the event (1-1500). The described system serves since 2009 and 
recorded a large data set of 194$\,\mathrm{M}$ physics quality collision events. 
Data taking is foreseen through at least 2019, with possible 
upgrades to the ToF detector.

\section*{Acknowledgments}

This work was supported in part by the Momentum (``Lend\"ulet'') program of the 
Hungarian Academy of Sciences, as well as the J\'anos Bolyai Research 
Scholarship of the Hungarian Academy of Sciences. Funding from the grant 
NKTH 68506 is also acknowledged. 
We wish to thank to the support and help from all the members of the CERN 
NA61/SHINE Collaboration.



\section*{References}

\begin{thebibliography}{99}


\bibitem{na61homepage} NA61 webpage: \texttt{http://na61.web.cern.ch}
\bibitem{na61proposal} N.~Abgrall \emph{et al} (the NA61 Collaboration): \emph{Study of Hadron Production in Hadron-Nucleus and Nucleus-Nucleus Collisions at the CERN SPS}; Proposal for the NA61/SHINE experiment [\texttt{http://cdsweb.cern.ch/record/995681/files/ spsc-2006-034.pdf}].
\bibitem{na61cpod07}   A.~L\'aszl\'o (for the NA61 Collaboration): \emph{NA61/SHINE at the CERN SPS}; PoS \textbf{CPOD07} (2007) 054.
\bibitem{na61qm09}     A.~L\'aszl\'o (for the NA61 Collaboration): \emph{The NA61/SHINE Experiment at the CERN SPS}; Nucl.~Phys.~\textbf{A830} (2009) 559c.
\bibitem{na61nim}      N.~Abgrall \emph{et al} (the NA61 Collaboration): \emph{NA61/SHINE facility at the CERN SPS: beams and detector system}; J.~Instr.~\textbf{9} (2014) P06005.
\bibitem{na49detector} S.~Afanasiev \emph{et al} (the NA49 Collaboration): \emph{The NA49 Large Acceptance Hadron Detector}; Nucl.~Instr.~Meth.~\textbf{A430} (1999) 210.
\bibitem{lmpd}         K.~M\'arton, G.~Kiss, A.~L\'aszl\'o, D.~Varga: \emph{Low momentum particle detector for the NA61 experiment at CERN}; Nucl.~Instr.~Meth.~\textbf{A763} (2014) 372.
\bibitem{na49readout}  F.~Bieser \emph{et al}: \emph{Design and performance of TPC readout electronics for the NA49 experiment}; Nucl.~Instr.~Meth.~\textbf{A385} (1997) 535.
\bibitem{na49fesca}    S.~A.~Kleinfelder: \emph{A 4096 Cell Switched Capacitor Analog Waveform Storage Integrated Circuit}; IEEE~Transactions~on~Nucl.~Sci.~\textbf{37} (1990) 1230.
\bibitem{ddl}          G.~Rubin, P.~Vande~Vyvre, P.~Csat\'o, E.~D\'enes, T.~Kiss, Z.~Meggyesi, J.~Sulyan, G.~Vesztergombi, B.~Eged, I.~Gerencs\'er, I.~Nov\'ak, C.~So\'os, D.~Tarj\'an, A.~Telegdy, N.~T\'oth: \emph{The ALICE Detector Data Link}; Proceedings of the Fifth Workshop on Electronics for LHC Experiments (LECC 1999), p.493.
\bibitem{rorc}         W.~Carena, F.~Carena, S.~Chapeland, E.~D\'enes, R.~Divi, T.~Kiss, J.~C.~Marin, K.~Schossmaier, C.~So\'os, A.~Vascotto, P.~Vande~Vyvre: \emph{The ALICE Data-Acquisition Read-out Receiver Card}; Proceedings of the 10th Workshop on Electronics for LHC Experiments (LECC 2004), p.273.
\bibitem{specs}        E.~D\'enes, T.~T\"olyhi, C.~So\'os, T.~Kiss, Z.~Fodor, A.~L\'aszl\'o: \emph{NA61 TPC Read-out}; Specification of the NA61 TPC read-out electronics and firmware (2010), NA61 Internal Note.
\bibitem{psd}          M.~Golubeva \emph{et al}: \emph{Forward hadron calorimeter for measurements of projectile spectators in heavy-ion experiment}; Phys.~Atom.~Nucl.~\textbf{75} (2012) 673.
\bibitem{drs}          S.~Ritt \emph{et al}: \emph{Application of the DRS chip for fast waveform digitizing}; Nucl.~Instr.~Meth.~\textbf{A623} (2010) 486.
\bibitem{tcltk}        Tcl/Tk: \texttt{http://www.tcl.tk}
\bibitem{readline}     GNU Readline Library: \texttt{http://www.gnu.org/s/readline}
\bibitem{shine}        R.~Sipos, A.~L\'aszl\'o, A.~Marcinek, T.~Paul, M.~Szuba, M.~Unger, D.~Veberic,  O.~Wyszy\'{n}ski: \emph{The Offline Software Framework of the NA61/SHINE Experiment}; J.~Phys.~Conf.~Ser.~\textbf{396} (2012) 022045.
\bibitem{castor}       CASTOR: \texttt{http://castor.web.cern.ch}

\end{thebibliography}
\end{document}